\documentclass[prl,twocolumn,lengthcheck,superscriptaddress]{revtex4-1}

\usepackage{amsmath}
\usepackage{amsthm}
\usepackage{amssymb}
\usepackage{mathrsfs}
\usepackage{graphicx}
\usepackage{hyperref}
\usepackage{color}
\usepackage{braket}

\newcommand{\tr}{\operatorname{tr}}

\newcommand{\supp}{\operatorname{supp}}
\newcommand{\dist}{\operatorname{dist}}

\theoremstyle{definition}

\theoremstyle{remark}

\begin{document}

\title{Continuum Limits of Quantum Lattice Systems}

\author{Tobias J.\ Osborne}
\affiliation{Institut f\"ur Theoretische Physik, Leibniz Universit\"at Hannover, Appelstr. 2, 30167 Hannover, Germany}

\begin{abstract}
	We describe a general procedure to give effective continuous descriptions of quantum lattice systems in terms of quantum fields. There are two key novelties of our method: firstly, it is framed in the hamiltonian setting and applies equally to distinguishable quantum spins, bosons, and fermions and, secondly, it works for arbitrary variational tensor network states and can easily produce computable non-gaussian quantum field states. Our construction extends the mean-field fluctuation formalism of Hepp and Lieb (developed later by Verbeure and coworkers) to identify \emph{emergent} continuous large-scale degrees of freedom --- the continuous degrees of freedom are not identified beforehand. We apply the construction to to tensor network states, including, matrix product states and projected entangled-pair states, where we recover their recently introduced continuous counterparts, and also for tree tensor networks and the multi-scale entanglement renormalisation ansatz. Finally, extending the continuum limit to include dynamics we obtain a strict light cone for the propagation of information. 
\end{abstract}

\maketitle

The modern understanding of quantum field theory (QFT), emphasised by Wilson \cite{wilson:1975a}, is as an \emph{effective theory} describing large lengthscale physics. Thus it has become a powerful tool for the description of the behaviour of complex quantum systems, including, critical phenomena \cite{sachdev:2011a} and topological order \cite{wen:2007a}. Quantum-field reasoning is especially powerful when the system is effectively modelled by a weakly interacting QFT, enabling the full power of perturbative methods to be unleashed. Unfortunately, however, the QFT approach is not a universal panacea: when one obtains a strongly coupled effective QFT description, e.g., a nonabelian gauge theory, then perturbative methods are not so easily applied and one must resort to numerical methods, largely obviating the simplifications afforded by the effective continuous description \footnote{Note that there are very powerful general methods based on the supersymmetric method \cite{efetov:1997a}, and yet others on the AdS/CFT correspondence \cite{sachdev:2010a}, that can be exploited in the nonperturbative regime.}. In such cases a \emph{tensor network} \cite{orus:2013a,bridgemanHandwavingInterpretiveDance2017} may afford a more parsimonious representation for the physics of the system.

Tensor network states (TNS) have enjoyed, thanks to insights from quantum information theory in the study of entanglement, a recent renaissance allowing the analytic and numerical investigation of many intriguing strongly correlated phenomena, including, dynamics \cite{vidal:2003a, haegeman:2011b,osborne:2005d}, fermions \cite{corboz:2009a, corboz:2010a, corboz:2010b, kraus:2010a} without sign problems, the determination of spectral information \cite{haegeman:2012a}, topological order \cite{aguado:2008a, buerschaper:2009a, koenig:2009a}, and dissipative quantum phenomena \cite{verstraete:2004b, zwolak:2004a, kraus:2012a}. There are, however, many remaining challenges. One outstanding problem is that it is difficult to identify the correct effective continuous degrees of freedom describing the experimentally relevant long-range behaviour of a TNS \cite{verstraete:2010a, haegeman:2011a,jenningsContinuumTensorNetwork2015}. Thus it is desirable to develop a more systematic and operationally well-motivated procedure to obtain an effective continuous description. Such a method, if developed, would also potentially contribute to the study of nonperturbative QFT because the TNS ansatz could then be systematically lifted to the continuous setting potentially giving new computable nongaussian vacua. 

One standard way to obtain an effective QFT description of a system is based on the use of the path-integral representation (see, e.g., \cite{wilson:1975a, auerbach:1994a}): one first represents the partition function using the path integral. Effective large-lengthscale continuous degrees of freedom are then identified and the path integral approximated by a continuum field path integral. This procedure has been applied with considerable success to a tremendous variety of lattice systems (see, e.g., \cite{sachdev:2011a, fradkin:2013a}). However, this approach does not apply so readily to systems efficiently described in terms of tensor network states. Although there are now path-integral representations for several TNS classes \cite{brockt:2012a,jenningsContinuumTensorNetwork2015}, it is not clear how to use this representation to identify the \emph{relevant} degrees of freedom for such a network. It is almost certainly a red herring to exploit a stationary phase argument for the path integral representation here as this could easily identify auxiliary degrees of freedom which lead to trivial changes in the physical state. Indeed, the question of identifying the operationally relevant degrees of freedom for the description of the large-scale features of a complex quantum system has only recently been initiated \cite{beny:2013a,benyInferringEffectiveField2017a}.   

In this Letter we describe a construction which produces an effective continuous description of a quantum lattice system whether it be in a TNS or a gaussian state. The first step is to identify a classical mean-field like continuous description, which we do by exploiting a generalised mean-field approach along the lines of \cite{verbeure:2011a} and \cite{raggio:1989a,raggio:1991a}. The operationally significant quantum fluctuations around this classical state are then identified and represented in terms of a family quantum fields in a systematic way. We achieve this by generalising the quantum fluctuation algebra construction originated by Hepp and Lieb \cite{hepp:1973b} (later generalised by Verbeure and coworkers  \cite{goderis:1989d,goderis:1989c,goderis:1990b,goderis:1989b,goderis:1990a,goderis:1988a,goderis:1989a,michoel:1998a,requardt:2002a, verbeure:2011a}) to produce a quantum field fluctuation algebra. The result of this construction is a list of quantum-field data for the system: (i) a quantum field ``ground state'' $|\Omega\rangle$; and (ii) a list of quantum fields for the fluctuations with a precise map to the corresponding lattice quantities. This construction may be applied to matrix product states, projected entangled-pair states, and tree-type tensor networks such as the multiscale entanglement renormalization ansatz. (When applied to quasi-free lattice systems such as the lattice Klein-Gordon equation, one recovers the usual QFT continuum limit.) Extending the continuum limit to apply to dynamics we obtain strict causality: information may not propagate outside of a light-cone. 

\paragraph{Preliminaries}\hspace{-1em}.---Our construction applies to systems arranged on a lattice $a\mathbb{Z}^D$ embedded in $\mathbb{R}^D$, where $a>0$ is the lattice spacing. For simplicity we mostly restrict to one-dimensional quantum spin systems with a single distinguishable quantum spin $\mathbb{C}^d$ attached to each site. (These assumptions entail no essential restrictions: all our subsequent constructions apply immediately to higher-dimensional systems and even to irregular lattices embedded in $\mathbb{R}^D$.) Thus the Hilbert space for our system is morally associated with the infinite tensor product $\bigotimes_{j\in\mathbb{Z}} \mathbb{C}^d$. Considerable care should be exercised when working with such tensor products; the safest course of action, and the one adopted here, is to work with the space $\mathcal{A}(\mathbb{Z})$ of observables of the spin system instead \footnote{Let $d\in \mathbb{N}$ and $x \in \mathbb{Z}^D$ and write $\mathcal{A}_x  \cong M_d(\mathbb{C})$, where $M_d(\mathbb{C})$ is the algebra of $d\times d$ complex matrices. Then for any finite subset $\Lambda \subset \mathbb{Z}^D$ let $\mathcal{A}(\Lambda)$ be the tensor product of $\mathcal{A}_x$ over all $x\in \Lambda$. For $\Lambda_1\subset \Lambda_2$ identify $\mathcal{A}(\Lambda_1)$ with the subalgebra $\mathcal{A}(\Lambda_1)\otimes \mathbb{I}_{\Lambda_2\setminus \Lambda_1} \subset \mathcal{A}(\Lambda_2)$. For infinite $\Lambda \subset \mathbb{Z}^D$ denote by $\mathcal{A}(\Lambda)$ the $C^*$-closure of the increasing family of finite-dimensional algebras $\mathcal{A}(\Lambda_f)$ with $\Lambda_f \subset \Lambda$. The \emph{quasi-local algebra} is then $\mathcal{A}(\mathbb{Z}^D)$.}.

Let $j\in\mathbb{Z}$ and $A\in \mathcal{A}\equiv M_d(\mathbb{C})$. Then we usually write $A_j$ for the observable $A$ localised at spin $j$: $A_j \equiv \mathbb{I}_{\cdots j-2, j-1}\otimes A_j\otimes \mathbb{I}_{j+1,j+2,\cdots}$. We often require a basis for $\mathcal{A}$: to this end let $\lambda^{\alpha}$ be an orthonormal hermitian operator basis for the single-site observables $\mathcal{A}$ (according to the Hilbert-Schmidt inner product).

\begin{figure}
\includegraphics{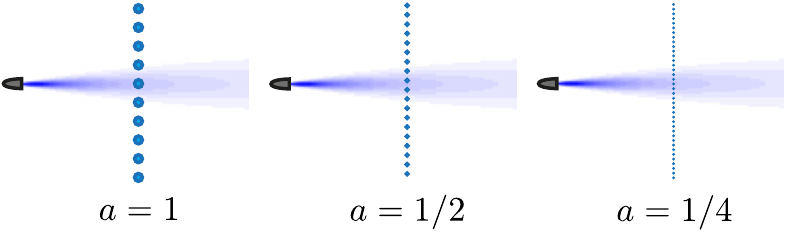}
\caption{An illustration of the continuum limit: measuring the observable $\phi^{(a)}( f_x)$. Here the quantum lattice is illustrated for several different values of the lattice spacing $a$. As $a$ gets smaller any fixed measurement of the quantum spin centred on $x$ will necessarily address more and more spins. The continuum limit is given by a limiting value $\phi(f_x)$}
\end{figure}
\paragraph{The classical continuum limit}\hspace{-1em}.---The key to understanding how to take a continuum limit is to work out an operationally meaningful way to compare states $\rho_a$ and $\rho_{a'}$ associated with \emph{different} lattice spacings $a\not=a'$. Naively this is impossible unless $a=a'$ because the quantum spins typically live at different locations. The crucial idea here is to compare states not directly by the fidelity but by instead agreeing on a family of \emph{observables} $\{O_a\}_{a\in\mathbb{R}^+}$, indexed by $a$, which are understood to refer to the \emph{same} experiment, only discretised on a lattice with spacing $a$. If $\rho_a(O_a) \approx \rho_{a'}(O_{a'})$ for all observables $O_a$ then we declare the two states to be close. A physical justification for this notion is as follows. Imagine we have a quantum spin chain and we perform a neutron-type scattering experiment (see Fig.~1 for an illustration) to measure, for example, the $z$-component of a spin at physical position $x$. The impinging beam of neutrons, even if well collimated, will inevitably spread as it travels toward the spin chain. Thus the observable measured by this scattering experiment is, instead of $\sigma^z_{\lfloor x/a \rfloor}$, rather something like $\phi^{(a)}(f_x) \equiv a\sum_{j\in\mathbb{Z}} f_x(aj) \sigma_j^z$, where $f_x$ is the \emph{beam shape} (for example, a gaussian centred at $x$: $f_x(y) \approx e^{-(y-x)^2}$). Thus the beam addresses approximately $1/a$ spins. Even though the lattice spacing $a$ of the quantum lattice system isn't precisely known, we declare that $\phi^{(a)}(f_x)$ corresponds the same experimental setup, i.e., the ``measurement of the spin at location $x$''. 

Motivated by this example we declare the family of observables given by
\begin{equation}
	\phi^{(a)}(f) \equiv a\sum_{j\in\mathbb{Z}} \sum_{\alpha=0}^{d^2-1} [f]_\alpha(aj) \lambda_j^\alpha,
\end{equation}
where $f$ is a $d^2$-dimensional vector-valued function \footnote{Suppose that $v$ is a $d$ dimensional vector. We exploit the notation $[v]_\alpha$, $\alpha=0,1,\ldots, d-1$ for the components of the vector.}, and all products and linear combinations thereof, to be the correct method to compare states on quantum lattices with different lattice spacings. That is, we say $\rho_a$ and $\rho_{a'}$ are close if $\tr(\rho_a \phi^{(a)}(f_1)\cdots \phi^{(a)}(f_m)) \approx \tr(\rho_{a'} \phi^{(a')}(f_1)\cdots \phi^{(a')}(f_m))$ for all possible choices of local rapidly decaying continuous vector-valued functions $f_l$, $l= 1, 2, \ldots, m$, with components $[f_l]_{\alpha}(x)$, for $\alpha = 0, 1, \ldots d^2-1$. 

The terminology ``classical continuum limit'' is justified upon noting the important fact that the operators $\phi^{(a)}(f)$ commute with each other in the limit $a\rightarrow 0$: $\lim_{a\rightarrow 0}\|[\phi^{(a)}(f), \phi^{(a)}(g)]\| =  0$. Thus, in the limit $a\rightarrow 0$, the observables $\phi^{(a)}(f)$ may be jointly measured; the set of limiting observables forms a commutative algebra and therefore models an effective classical system \footnote{If, instead, we worked with a lattice of fermions instead of quantum spins we would see that the operators $\phi_{A}^{(a)}(f)$ \emph{anticommute} with each other in the limit $a\rightarrow 0$, i.e., they give a representation of the grassmann numbers.}.

The classical continuum limit is then \emph{defined} by the expectation values of $\phi^{(a)}(f)$ for $a\rightarrow 0$, as $f$ runs over all possible ``beam shapes'':
\begin{equation}
	\left\langle\prod_{l=1}^m\phi(f_l)\right\rangle \equiv \lim_{a\rightarrow 0} \tr\left(\rho_a \prod_{l=1}^m\phi^{(a)}(f_l)\right),
\end{equation}
provided this limit exists. We say that $\{\rho_a\}_{a\in\mathbb{R}^+}$ \emph{admits a classical continuum limit} if these expectation values are finite for all rapidly decaying vector-valued functions $f_l$ \footnote{By ``rapidly decaying function'' we mean \emph{tempered distribution}.}. Note the relation $\phi(f+g) = \phi(f) + \phi(g)$.

It is worth pausing to explore a couple of simple examples to get a feeling for the classical continuum limit construction. Consider the case where $\mathcal{A} = M_2(\mathbb{C})$, i.e., our system is a chain of qubits, and $\rho_a = \bigotimes_{j\in\mathbb{Z}} \varrho$, where $\varrho$ is some single-qubit state. That is, $\rho_a$ is a translation-invariant mean-field ansatz independent of lattice spacing. In this case it is relatively easy to show that $\left\langle\prod_{l=1}^m\phi(f_l)\right\rangle = \prod_{l=1}^m\left\langle\phi(f_l)\right\rangle$ and $\left\langle\phi(f)\right\rangle = \sum_{\alpha=0}^3\tr(\lambda^\alpha\varrho) \int [f]_\alpha(x) \, dx$.

The second example is hardly different from the first, except that we now we relax translation invariance and set $\rho_a = \bigotimes_{j\in\mathbb{Z}} \varrho(ja)$, with $\varrho(x) \equiv \frac{1}{2}\sum_{\alpha=0}^3r_{\alpha}(x)\sigma^\alpha$, where $r_\alpha(x) = (1, r_1(x), r_2(x), r_3(x))$  is some continuous vector-valued function of $x\in \mathbb{R}$ and $\sigma^0 \equiv \mathbb{I}$, $\sigma^1 \equiv \left(\begin{smallmatrix}0 & 1\\ 1 & 0\end{smallmatrix}\right)$, $\sigma^2 \equiv  \left(\begin{smallmatrix}0 & -i\\ i & 0\end{smallmatrix}\right)$, and  $\sigma^3 \equiv \left(\begin{smallmatrix}1 & 0\\0 & -1\end{smallmatrix}\right)$. In this case we again find that $\left\langle\prod_{l=1}^m\phi(f_l)\right\rangle = \prod_{l=1}^m\left\langle\phi(f_l)\right\rangle$. However, we now have that $\left\langle\phi(f)\right\rangle =  \sum_{\alpha=0}^3\int [f]_\alpha(x) r_\alpha(x) \, dx$.

We now implicitly define the (vector) \emph{field operator} $\phi(x)$ to be that object which gives $\phi(f)$ according to the relation 
\begin{equation}
	\phi(f) = \sum_{\alpha=0}^{d^2-1}\int f_\alpha(x) [\phi]_\alpha(x)\,dx, 
\end{equation}
for all smearing funcitons $f$. For the first example above we find that  $\phi(x)$ takes a constant definite value: $[\phi]_\alpha(x) = \tr(\sigma^\alpha \varrho)$. In the second example the field takes a nonconstant definite value: $[\phi]_\alpha(x) = r_\alpha(x)$, with $r_0(x) = 1$.  In both cases the observable $\phi(x)$ takes some definite value.

However, the field operator $\phi(x)$ is not a simple function. We emphasise that $\phi(x)$ is a classical \emph{observable}, i.e.\ a random variable, so that in the case where $\rho_a$  yields a statistical ensemble in the classical continuum limit --- an example is $\rho_a = \frac12 \rho_a' + \frac12 \rho_a''$, with $\rho_a'$ and $\rho_a''$ taking different classical continuum limits --- we see that $\phi(x)$ doesn't always take a definite value. 

\paragraph{Fluctuations: the quantum continuum limit}\hspace{-1em}.---The classical continuum limit construction is essentially a law of large numbers result: the operators $\phi^{(a)}$ are the (weighted) arithmetic mean of roughly $1/a$ independent observables so that if the sequence $\{\rho_a\}_{a\in\mathbb{R}^+}$ is sufficiently well behaved we are guaranteed the existence of the limit. To see any quantum structure in the continuum we need to analyse and model the \emph{fluctuations} around the mean value of the observables $\phi^{(a)}(f)$ as the lattice spacing is decreased. Such fluctuations are detected by observables of the form
\begin{equation}
	\widetilde{\phi}^{(a)}(f) \equiv Z_f(a)\left( a\sum_{\alpha=0}^{d^2-1}\sum_{j\in \mathbb{Z}} [f]_\alpha(aj) \lambda^\alpha_j - \langle\phi(f)\rangle\mathbb{I}\right),
\end{equation}
where $Z_f(a)$ is a \emph{field-strength renormalization} chosen to amplify the fluctuations. Central limit theorem considerations strongly suggests that these fluctuations will typically be on the order of $\sqrt{a}$; so we tentatively choose $Z_f(a) = 1/\sqrt{a}$ in order to ensure the fluctuations remain present in the limit $a\rightarrow 0$. From the perspective of a scattering experiment interpretation what we are doing is increasing the sensitivity of our detection apparatus to amplify the fluctuations due to microscopic features: if the experimentalist leaves the sensitivity of the detector constant then this experiment will only be sensitive to the bulk classical fields $\phi(x)$ and the experimentalist is content with a classical model for the statistical fluctuations in $\phi(x)$ due to the ensemble of classical limits. If, however, the experimentalist improves the detector to be sensitive to fluctuations of the order of $\sqrt{a}$ then, on top of the original statistical fluctuations, \emph{additional} fluctuations will emerge, now with a possibly quantum explanation. We now make this precise. 

Analogous to the classical case, the quantum continuum limit is defined by the expectation values of $\widetilde{\phi}^{(a)}(f)$ for $a\rightarrow 0$, as $f$ runs over all rapidly decaying vector-valued functions:
\begin{equation}\label{eq:qctslimit}
	\langle\Omega|\left(\prod_{l=1}^m\widehat{\phi}(f_l)\right)|\Omega\rangle \equiv \lim_{a\rightarrow 0} \tr\left(\rho_a \prod_{l=1}^m\widetilde{\phi}^{(a)}(f_l)\right),
\end{equation}
provided this limit exists \footnote{Here the order of the products matters; we henceforth take all product symbols as running from left to right: $\prod_{l=1}^m M_l \equiv M_1M_2\cdots M_m$.}. Thus $\{\rho_a\}_{a\in\mathbb{R}^+}$ is said to admit a \emph{quantum continuum limit} if these expectation values are finite for all rapidly decaying functions $f_l$. It is sometimes necessary in the sequel to relax this requirement and only demand that a subset of the limits exist, typically for operators admitting an interpretation as canonical field amplitude observables (in such a case we obtain a \emph{singular} state, in the sense of operator algebras). This innocuous-seeming condition is deeply nontrivial; one of the main contributions of this paper is to show that there are families of tensor network states for which this condition is fulfilled. 

The expectation value in Eq.~(\ref{eq:qctslimit}) is taken with respect to a \emph{pure state} $|\Omega\rangle$; we now show how this state is constructed and justify the notation $\langle\Omega|\cdot |\Omega\rangle$ for the limiting functional. Because $\lim_{a\rightarrow0} \tr(\rho_a) = 1$ we have that the limiting functional is normalised: $\langle\Omega| \mathbb{I} |\Omega\rangle = 1$. Secondly, we inherit linearity in the limit: $\langle \Omega| (\alpha \widehat{\phi}(f) + \beta\widehat{\phi}(g))|\Omega\rangle = \lim_{a\rightarrow 0} \tr\left(\rho_a \left(\alpha \widetilde{\phi}^{(a)}(f) + \beta \widetilde{\phi}^{(a)}(g)\right) \right)$ $= \alpha\lim_{a\rightarrow 0} \tr\left(\rho_a  \widetilde{\phi}^{(a)}(f)\right)$ + $\beta\lim_{a\rightarrow 0} \tr\left(\rho_a \widetilde{\phi}^{(a)}(g)) \right)$ $=  \alpha \langle \Omega| \widehat{\phi}(f) |\Omega\rangle+ \beta\langle \Omega|\widehat{\phi}(g)|\Omega\rangle$. Finally we have that the limiting functional is positive: $\langle\Omega | \widehat{\phi}(f)^\dag \widehat{\phi}(f) |\Omega\rangle$ $= \lim_{a\rightarrow 0} \tr(\rho_a (\widetilde{\phi}^{(a)}(f)^\dag \widetilde{\phi}^{(a)}(f))) \ge 0$. (These arguments remain true for products of the fluctuation operators.) Thus the limiting object is a state. Further, we can always purify the limiting state to $|\Omega\rangle$, justifying the notation $\langle \Omega|\cdot|\Omega\rangle$.

Using the state $|\Omega\rangle$ we now describe the Hilbert space $\mathcal{H}$ of quantum field fluctuations: this is simply the Hilbert space built from $|\Omega\rangle$ and all vectors of the form $\left(\prod_{l=1}^m\widehat{\phi}(f_l)\right)|\Omega\rangle$. (It is a nontrivial fact that this construction leads to a separable Hilbert space.) In terms of these vectors a fluctuation operator $\widehat{\phi}(f)$ acts as
\begin{equation}	
	\widehat{\phi}(f) \left(\prod_{l=1}^m\widehat{\phi}(f_l)\right)|\Omega\rangle \equiv \widehat{\phi}(f) \widehat{\phi}(f_1)\cdots \widehat{\phi}(f_m)|\Omega\rangle.
\end{equation}

A remarkable consequence --- similar to the original quantum fluctuation construction \cite{hepp:1973b, verbeure:2011a} --- of the fluctuation field construction is that the limiting objects obey the bosonic canonical commutation relations: as long as $Z_f(a)Z_g(a) = 1/a$ we have that $[\widehat{\phi}(f), \widehat{\phi}(g)] = \langle \phi(f\wedge g)\rangle \mathbb{I}$ \footnote{The wedge product between two vector-valued functions $f$ and $g$ is defined as follows. First write $\lambda^\alpha\lambda^\beta =  \frac12{d^{\alpha\beta}}_\gamma \lambda^\gamma + \frac{i}{2}{f^{\alpha\beta}}_{\gamma}\lambda^\gamma$ for the structure constants of the operator basis $\lambda^\alpha$, where ${d^{\alpha\beta}}_\gamma$ (respectively, ${f^{\alpha\beta}}_{\gamma}$) are the completely symmetric (respectively, antisymmetric) structure constants. The wedge product of two vectors $u$ and $v$  is then defined to be the vector with components $[u\wedge v]_{\gamma} = \sum_{\alpha,\beta=0}^{d^2 -1}[u]_\alpha[v]_\beta {f^{\alpha\beta}}_\gamma$. Thus the wedge product $f\wedge g$ of two vector-valued functions is the vector-valued function $f(x)\wedge g(x)$.}. This surprising fact is proved as follows.  Consider
\begin{equation}
	\begin{split}
		[\widetilde{\phi}^{(a)}(f), \widetilde{\phi}^{(a)}(g)] &= a^2Z_f(a)Z_g(a)\sum_{j\in\mathbb{Z}} [f\wedge g]_\gamma(aj) \lambda^\gamma_j \\
		&= \phi^{(a)}(f\wedge g).
	\end{split}
\end{equation} 
This expression may be rewritten as a local fluctuation operator:
\begin{equation}
	[\widetilde{\phi}^{(a)}(f), \widetilde{\phi}^{(a)}(g)] = \langle\phi(f\wedge g)\rangle \mathbb{I} + Z_{f\wedge g}^{-1}(a)\widetilde{\phi}^{(a)}(f\wedge g).
\end{equation}
Assuming that ${Z_{f\wedge g}^{-1}(a)} \underset{a\rightarrow 0}{\longrightarrow} 0$ we have that $[\widetilde{\phi}^{(a)}(f), \widetilde{\phi}^{(a)}(g)]$ is given by a constant term plus an operator decaying to $0$ on the hilbert space of fluctuations. Thus, in any expectation value, we can replace any commutator $[\widehat{\phi}(f), \widehat{\phi}(g)]$ with the number $\langle \phi(f\wedge g)\rangle \mathbb{I}$, i.e., 
\begin{equation}\label{eq:ccr}
	[\widehat{\phi}(f), \widehat{\phi}(g)] = \langle \phi(f\wedge g)\rangle \mathbb{I}.
\end{equation}
We now identify this algebraic structure with the canonical commutation relations. 

To understand the algebraic properties of the operators $\widehat{\phi}(f)$ it is convenient (but not necessary) to specialise to the translation invariant case. Here we have that 
\begin{equation}
	\langle \phi(f\wedge g)\rangle = \lim_{a\rightarrow 0} \tr(\rho_a [\lambda^\alpha,\lambda^\beta]) ([\overline{f}]_\alpha,[g]_\beta), 
\end{equation}
where $(f,g) = \int \overline{f}(x)g(x)\,dx$.
Form the hermitian antisymmetric matrix $\Theta$ with matrix elements $[\Theta]_{\alpha\beta}  \equiv \lim_{a\rightarrow 0} \tr(\rho_a [\lambda^\alpha,\lambda^\beta])$. The matrix $\Theta$ induces an antisymmetric form $\Theta(A,B) \equiv \sum_{\alpha,\beta} a_\alpha b_\beta [\Theta]_{\alpha\beta}$ for $A = \sum_{\alpha=0}^{d^2-1} a_\alpha \lambda^\alpha$ and  $B = \sum_{\alpha=0}^{d^2-1} b_\beta \lambda^\beta$. Applying the symplectic Gram-Schmidt procedure to this symplectic form produces a list of three hermitian operators $\{\mu^{j}, \nu^{j}\}_{j=1}^m$ and  $\{\xi^{j}\}_{j = 1}^{D^2-2m}$ such that $\Theta(\mu^j, \nu^k) = i\delta^{jk}$ and $\Theta(\xi^j, M) = 0$ for any $M$ which is a linear combination of $\mu^j$ and $\nu^j$. Let $f$ be a function with rapid decay and define $[a_j]_\alpha(x) \equiv f(x) \tr(\lambda^\alpha\mu^j)$, $[b_j]_\alpha(x) \equiv f(x) \tr(\lambda^\alpha\nu^j)$, and $[c_j]_\alpha(x) \equiv f(x) \tr(\lambda^\alpha\xi^j)$ We then define $\widehat{\varphi}_j(f) \equiv \widehat{\phi}(a_j)$ and $\widehat{\pi}_j(f) \equiv \widehat{\phi}(b_j)$, and $\widehat{\xi}_j(f) \equiv \widehat{\phi}(c_j)$. According to Eq.~(\ref{eq:ccr}) we then have that
\begin{equation}
	[\widehat{\varphi}_j(f), \widehat{\pi}_k(g)] = i (\overline{f},g)\delta^{jk} \mathbb{I}.
\end{equation}
Both of the operators $\widehat{\varphi}_j$ and $\widehat{\pi}_j$ are hermitian.
Finally, we implicitly define the quantum field operator $\widehat{\varphi}_j(x)$ via
\begin{equation}
	\widehat{\varphi}_j(f) \equiv \int f(x)\widehat{\varphi}_j(x)\,dx,
\end{equation}
and similarly for $\widehat{\pi}_j(x)$. In the case where the original lattice system was fermionic we obtain the canonical anticommutation relations instead.

Let's now apply the quantum continuum limit construction to some examples. The first is the product-state example $\rho_a = \bigotimes_{j\in\mathbb{Z}}\varrho$. We first construct the matrix $\Theta(\sigma^\alpha,\sigma^\beta) \equiv \frac12\sum_{\gamma=0}^3 r_\gamma \tr([\sigma^\alpha,\sigma^\beta]\sigma^\gamma)$. We assume, for simplicity, that $r = (0,0,0,1)$. (The general case is discussed in the Appendix.) In this case we find that $\mu^0 = \mathbb{I}$, $\mu^1 = \sigma^z$, $\nu^1 = \sigma^x$, and $\xi^1 = \sigma^y$. We thus recover standard fock space with vacuum state $|\Omega\rangle$.

The second example we study is that of a sequence of matrix product states $|\psi_a\rangle = \sum\langle \omega_L| \cdots A^{z_{-1}}_aA^{z_{0}}_a A^{z_{1}}_a\cdots |\omega_R\rangle |\cdots z_{-1}z_0z_{1} \cdots \rangle$
with the prescription $A^0_a \equiv \mathbb{I} + a Q$ and $A^1_a \equiv \sqrt{a} R$, where $Q$ and $R$ are $D\times D$ complex matrices. In this case we again find that the classical limit is a trivial product. The theta matrix is identical to the product state case, and we obtain the same quantum fluctuation operators. Thus the quantum field modelling the fluctuations around an MPS is precisely that of a single bosonic degree of freedom. However, in contrast to the product state case, the continuum limit state $|\Omega\rangle$ is \emph{not} the fock vacuum and nor is it generically gaussian. Indeed, it is precisely a continuous matrix product state $|\Omega\rangle \equiv \langle\omega_L|\mathcal{P}e^{\int_{-\infty}^\infty Q\otimes\mathbb{I} + R\otimes \widehat{\psi}^\dag(s)\,ds}|0\rangle|\omega_R\rangle$,  where $|0\rangle$ is the fock vacuum. 

Our third example concerns tree tensor network states. Here one obtains a continuum limit that coincides with a recently introduced limit investigated by Jones \cite{jonesScaleInvariantTransfer2017,jonesUnitaryRepresentationsThompson2014,jonesNogoTheoremContinuum2016,osborneDynamicsHolographicCodes2017a}. Further details of this example may be found in the Supplementary Material.

\paragraph{Dynamics of quantum fluctuations and strict causality}\hspace{-1em}.---Suppose we have a sequence $\rho_a$ giving rise to a quantum continuum limit and suppose, further, that we have a sequence  of hamiltonians $H_a\ge 0$ such that $\tr(H_a\rho_a) = 0$, $\forall a$. We can study the dynamics $U_a(t) = e^{itH_a}$ this sequence of hamiltonians generates for fluctuations around the continuum limit as follows. If $H_a$ is a sequence of nearest-neighbour hamiltonians for a quantum spin system then simple dimensional analysis requires that we must scale time via $t\mapsto t/a$. Thus we study the dynamics generated by the unitary $U_a(t/a)$ in the continuum limit. It is natural to expect that there is a continuum hamiltonian $\widehat{H}$ which generates the limiting dynamics. A useful mnemonic to derive this hamiltonian is that we can identify $\lambda^{\alpha}_{j} \equiv {Z_{\lambda^\alpha}^{-1}}\widetilde{\phi}^{(a)}(\delta_{\alpha\beta}\delta(ja))$. 

Rather astonishingly it turns out (see the Supplementary Material for details) that the dynamics generated by $\widehat{H}$ obeys strict causality: there is an \emph{exact} lightcone for the propagation of information through the continuous system. This is a consequence of the Lieb-Robinson bound  \cite{lieb:1972a,bratteli:1997a,nachtergaele:2010b}: we have that there is a universal constant $c$ for all $A$ and $B$ such that
\begin{equation}
	[\widehat{\phi}(f), e^{-it\widehat{H}} \widehat{\phi}(g)e^{it\widehat{H}} ] = 0
\end{equation}
whenever $\dist(\supp(f) , \supp(g)) < ct$. However, we don't generally obtain Poincar\'e invariance of the quantum continuum limit unless the excitation structure of $H$ exhibits more structure (rotation invariance is typically broken in most lattice models --- an effect that persists in the continuum limit.) 

As a simple application of this prescription we study the continuum limit of the product states $\rho_a = \bigotimes |0\rangle\langle 0|$ with corresponding hamiltonian given by the ferromagnetic $XY$ model, i.e., $H_a = \frac{1}{a}\sum_{j} \mathbb{I} - \sigma^x_j\sigma_{j+1}^x - \sigma^y_j\sigma_{j+1}^y$. Following the identification $\sigma^x_j \equiv \sqrt{a}\widetilde{\varphi}^{(a)}(\delta(ja))$, $\sigma^y_j \equiv \sqrt{a}\widetilde{\pi}^{(a)}(\delta(ja))$, and $\sigma^z_j \equiv \sqrt{a}\widetilde{\xi}^{(a)}(\delta(ja))$ we find
\begin{equation}
	\widehat{H} = \int \widehat{\varphi}(x) \partial_x\widehat{\varphi}(x) + \widehat{\pi}(x) \partial_x\widehat{\pi}(x) \, dx.
\end{equation}
We thus recover the result that, in the heisenberg picture, the fluctuation field obeys the wave equation:
\begin{equation}
	\partial_{tt}\widehat{\varphi}(x,t) = \partial_{xx}\widehat{\varphi}(x,t).
\end{equation}
Solutions to this equation obey strict causality, in accordance with the argument above.

\paragraph*{Conclusions}\hspace{-1em}.---We have introduced a procedure, based on the \emph{fluctuation construction} of Hepp and Lieb, Verbeure and coworkers, which takes a sequence of states of a quantum lattice system and produces a corresponding list of quantum field data for the limiting state. This method was applied to several natural tensor network classes yielding their well-known (and not so well-known) continuous analogues. It would be interesting to compare the continuum limit here with the projective construction of Kijowski \cite{kijowskiModificationProjectiveConstruction2017,okolowConstructionSpacesKinematic2013a}, developed further by Thiemann and coworkers \cite{laneryProjectiveLimitsState2017,laneryProjectiveLimitsState2017a,laneryProjectiveLimitsState2018,laneryProjectiveLimitsState2018a}, and also with the continuum limit described by Jones \cite{jonesScaleInvariantTransfer2017,jonesNogoTheoremContinuum2016,jonesUnitaryRepresentationsThompson2014} in the study of Thompson's groups $F$, $T$, and $V$. Indeed, for the case of Thompson's group $T$ it turns out the continuum limit described here coincides with the natural limit for fields (this will be the subject of a forthcoming paper), see also \cite{brothierOperatoralgebraicConstructionGauge2019} for recent progress in this direction in the context of $(1+1)$-dimensional gauge theories.

\acknowledgments{Numerous helpful discussions with C\'edric B\'eny, Marcus Cramer, Andrew Doherty, Terry Farrelly, Steve Flammia, Jutho Haegeman, Gerard Milburn, Terry Rudolph, Tom Stace, Frank Verstraete, Michael Walter, and Reinhard Werner are gratefully acknowledged. This work was supported in part by the ERC grants QFTCMPS and SIQS, the DFG through SFB 1227 (DQ-mat), the RTG 1991, and the cluster of excellence EXC 201 Quantum Engineering and Space-Time Research.}

\widetext
\appendix

\section{The spacetime continuum limit}
So far the continuum limit construction has been limited to the discussion of \emph{kinematics}, i.e., we only discussed the Hilbert-space structure for the continuous limit. In this appendix we extend the continuum limit construction to accommodate dynamics. 

The natural setting here is then that of a quantum lattice system in $D$ dimensions with hamiltonian 
\begin{equation}
	H_a \equiv \sum_{\langle j,k\rangle } h_{j,k}(a),
\end{equation}
where the sum is over all neighbouring spins. We allow the hamiltonian to possibly depend on the lattice spacing $a$, however, the norm $\|h_{j,k}(a)\|$ is required to be bounded by a constant for all sites $j$ and $k$ in the lattice. 

As we'll see later, such an identification leads to a definition of the dynamics for the continuum limit which obeys strict causality. This argument is essentially conditioned on the existence of a \emph{spacetime} quantum continuum limit which is defined in the following. First introduce the classical \emph{spacetime} discretised field operators via
\begin{equation}
	\phi^{(a)}(f) \equiv a^{D+1}\sum_{\alpha = 0}^{d^2-1}\sum_{j\in\mathbb{Z}^D} \int_{-\infty}^\infty f_\alpha(at, aj) e^{-itH_a}\lambda^\alpha_je^{itH_a} \, dt,
\end{equation}
where now $f$ is a rapidly decaying function on spacetime $\mathcal{M}_{D+1}\equiv\mathbb{R}\times \mathbb{R}^D$. We say that the sequence 
$\{\rho_a\}_{a\in\mathbb{R}^+}$ admits a \emph{classical spacetime continuum limit} if 
\begin{equation}
	\left\langle\prod_{l=1}^m\phi(f_l)\right\rangle \equiv \lim_{a\rightarrow 0} \tr\left(\rho_a \prod_{j=l}^m\phi^{(a)}(f_l)\right)
\end{equation}
for all rapidly decaying $f_l$ on $\mathcal{M}_{D+1}$.

We define the spacetime quantum fluctuation operators via
\begin{equation}
	\widetilde{\phi}^{(a)}(f) \equiv Z_f(a)\left( \int_{-\infty}^\infty  a^{D+1}\sum_{\alpha = 0}^{d^2-1}\sum_{j\in \mathbb{Z}^D} f_\alpha(at, aj) e^{-itH_a}\lambda^\alpha_je^{itH_a} \,dt - \langle\phi(f)\rangle\mathbb{I}\right).
\end{equation}
Correspondingly, the \emph{quantum spacetime continuum limit} is defined by the expectation values of products of $\widetilde{\phi}^{(a)}(f)$ for $a\rightarrow 0$, as $f$ runs over all rapidly decaying functions on $\mathcal{M}_{D+1}$:
\begin{equation}\label{eq:qsctslimit}
	\langle\Omega|\left(\prod_{l=1}^m\widehat{\phi}(f_l)\right)|\Omega\rangle \equiv \lim_{a\rightarrow 0} \tr\left(\rho_a \prod_{l=1}^m\widetilde{\phi}^{(a)}(f_l)\right),
\end{equation}
provided this limit exists.
(At this point it is worth noting that we can also obtain continuum limits for dynamics generated by discrete groups, e.g., quantum cellular automata, by replacing the integral over $t$ with a sum.)

\section{Strict causality for the quantum continuum limit}
In this appendix we argue that the dynamics of the quantum continuum limit obey strict causality. What we show is that
the continuum limit of any tensor network (or indeed, of any equilibrium state) with a local parent hamiltonian satisfies the following causality condition: there exists a constant velocity $c$ such that for all $f$ and $g$ are two functions on $\mathcal{M}_{D+1}$ whose supports are spacelike separated (with respect to the velocity $c$) we have that
\begin{equation}
	[\widehat{\phi}(f), \widehat{\phi}(g)]  = 0.
\end{equation}

To obtain this result we exploit the Lieb-Robinson bound (see, e.g., \cite{nachtergaele:2010b} and references therein)
\begin{equation}
	\| [A(t), B] \|_{\infty} \le C_0 \|A\|_\infty \|B\|_\infty e^{-{\tau} d(A,B) + \ell |t|}
\end{equation}
where
\begin{equation}
	A(t) \equiv e^{-itH_a} A e^{itH_a},
\end{equation} 
$d(A,B)$ is the number of edges in the shortest path through the lattice connecting the supports $\supp(A)$ and $\supp(B)$, and $C_0$, $\ell$, and $\tau$ are constants depending only $\sup \|h_{j,k}(a)\|_\infty$ and the maximum vertex degree of the lattice.

Consider 
\begin{equation}
	[\widetilde{\phi}^{(a)}(f), \widetilde{\phi}^{(a)}(g)] \equiv a^{2D+2} Z_f(a)Z_g(a) \sum_{\alpha,\beta=0}^{d^2-1}\sum_{j,k\in \mathbb{Z}^D}\int_{-\infty}^\infty\int_{-\infty}^\infty   f_\alpha(as, aj)g_\beta(at,ak) [e^{-isH_a}\lambda^\alpha_je^{isH_a}, e^{-itH_a}\lambda^\beta_ke^{itH_a} ] \,dsdt.
\end{equation}
We can bound this quantity according to
\begin{equation}
	\begin{split}
	\|[\widetilde{\phi}^{(a)}(f), \widetilde{\phi}^{(a)}(g)]\|_{\infty} &\le a^{2D+2} Z_f(a)Z_g(a) \sum_{\alpha,\beta=0}^{d^2-1}\sum_{j,k\in \mathbb{Z}^D}\int_{-\infty}^\infty\int_{-\infty}^\infty   |f_\alpha(as, aj)||g_\beta(at,ak)| \|[e^{-i(s-t)H_a}\lambda^\alpha_je^{i(s-t)H_a}, \lambda^\beta_k]\|_{\infty} \,dsdt \\
	&\le C_0 a^{2D+2} Z_f(a)Z_g(a)  \sum_{\alpha,\beta=0}^{d^2-1}\sum_{j,k\in \mathbb{Z}^D}\int_{-\infty}^\infty\int_{-\infty}^\infty   |f_\alpha(as, aj)||g_\beta(at,ak)| e^{-{\tau} d(j,k) + \ell |s-t|}\,dsdt,
	\end{split}
\end{equation}
where we've absorbed the constant $\max_{\alpha}\|\lambda^\alpha\|_\infty$ into $C_0$.
Changing variables we find
\begin{equation}
	\|[\widetilde{\phi}^{(a)}(f), \widetilde{\phi}^{(a)}(g)]\|_{\infty} \le C_0 d^4a^{2D} Z_f(a)Z_g(a)  \sum_{x_j,y_k\in a\mathbb{Z}^D}\int_{-\infty}^\infty\int_{-\infty}^\infty   f^\star(s', x_j)g^\star(t',y_k) e^{\frac{-{\tau} d(x_j,y_k) +\ell |s'-t'|}{a}}\,ds'dt',
\end{equation}
where
\begin{equation}
	f^\star(t,x) = \max_{\alpha} |f_\alpha(t,x)|, \quad g^\star(t,x) = \max_{\alpha} |g_\alpha(t,x)|,
\end{equation}
and
\begin{equation}
	d(x,y) = \sum_{\alpha=1}^D |x_\alpha -y_\alpha|
\end{equation}
is the $1$-norm distance between the vectors $x, y \in \mathbb{R}^D$.

Suppose now that the support of $f$ is spacelike separated with respect to the support of $g$, i.e., $\supp(g)$ lies outside the lightcone of $\supp(f)$. This means that 
\begin{equation}
		\inf_{\substack{x\in \supp(f)\\ y\in \supp(g)}} d(x,y) > c|s-t|,
\end{equation}
where $c = \ell/\tau$. Note that in our setting the lightcones are diamond shaped  (i.e., the ``speed of light'' varies with direction). This means that
\begin{equation}
	c|s-t| -d(x,y) < 0, \quad (s,x)\in\supp(f)\quad \text{and} \quad (t,y)\in \supp(g).
\end{equation}
that is,
\begin{equation}
	(e^{c|s-t| - d(x,y)})^{\frac{\tau}{a}} \le 1, \quad (s,x)\in\supp(f)\quad \text{and} \quad (t,y)\in \supp(g),
\end{equation}
Thus we conclude that
\begin{equation}
	\lim_{a\rightarrow 0} a^{-m}e^{\frac{-{\tau} d(x,y) +\ell |s-t|}{a}} = 0, \quad m\in \mathbb{Z}^+,\quad (s,x)\in\supp(f),\quad \text{and} \quad (t,y)\in \supp(g), 
\end{equation}
From this we have that
\begin{equation}
	\begin{split}
	\lim_{a\rightarrow 0}\|[\widetilde{\phi}^{(a)}(f), \widetilde{\phi}^{(a)}(g)]\|_{\infty} &\le  d^4C_0 \|f\|_\infty \|g\|_\infty\lim_{a\rightarrow 0}\int_{\supp(f)}\int_{\supp(g)}   a^{-m} e^{\frac{-{\tau} d(x,y) +\ell |s-t|}{a}}\,dsdxdtdy = 0.
	\end{split}
\end{equation}
where $m$ is determined by the largest integer such that $Z_f(a)Z_g(a) \le a^{-m}$.
Therefore we conclude that
\begin{equation}
	[\widehat{\phi}(f), \widehat{\phi}(g)]  = 0.
\end{equation}

Obviously the existence of a spacetime continuum limit is a rather demanding restriction, but there is one general setting where it is fulfilled, namely the case where $\rho_a$ is an equilibrium state for $H_a$, i.e., either the ground state or a thermal state. (More generally, we expect that the causality argument will work when $\{\rho_a(t)\}_{a\in \mathbb{R}^+}$ consists of a finite density of excitations on top of the equilibrium state.

\section{Examples}
\subsection{Product state case}
Here we consider the case where our sequence $\rho_a$ is a product, i.e., $\rho_a = \bigotimes_{j\in\mathbb{Z}} \varrho$, where $\varrho$ is some constant single-qubit state $\varrho \equiv \frac{1}{2}\sum_{\alpha=0}^3r_\alpha \sigma^\alpha$. Because the identity factorises out of the continuum limit it is somewhat convenient to adopt a $3$-vector notation where $r = (r^0, \mathbf{r})$ and  $\sigma = (\sigma^0, \boldsymbol{\sigma})$, with $\sigma^0 \equiv \mathbb{I}$ and $\boldsymbol{\sigma} \equiv \left[ \left(\begin{smallmatrix} 0 &1\\ 1& 0\end{smallmatrix}\right),  \left(\begin{smallmatrix} 0 &-i\\ i& 0\end{smallmatrix}\right),  \left(\begin{smallmatrix} 1 & 0\\ 0& -1\end{smallmatrix}\right) \right]$ so that $\varrho = \frac{1}{2}(\mathbb{I} + \mathbf{r}\cdot \boldsymbol{\sigma})$. We find for the $\Theta$ matrix (omitting the identity) the expression $[\Theta]_{jk} = \Theta(\sigma^j,\sigma^k) \equiv \frac12\sum_{l=1}^3 r_l \tr([\sigma^j,\sigma^k]\sigma^l) = i\sum_{l=1}^3 r_l \epsilon_{jkl}$, which gives
\begin{equation}
	\Theta = i\begin{pmatrix} 0 & r_z & -r_y \\  -r_z & 0 & r_x \\ r_y & -r_x & 0 \end{pmatrix}.
\end{equation}
We assume that $\|\mathbf{r}\| \not= 0$ (in the case where $\|\mathbf{r}\| = 0$ the continuum limit is trivial and there are no quantum fluctuations).
The symplectic Gram-Schmidt procedure produces the following three vectors (here we assume that $r_j \not= 0$, $j=1,2,3$; there are obvious modifications in the general case):
\begin{equation}
	\begin{split}
	e &= \frac{1}{\sqrt{r_x^2+r_y^2}\|\mathbf{r}\|}\begin{pmatrix} r_y \\ -r_x \\ 0\end{pmatrix}, \\  \quad f &= \frac{1}{\sqrt{r_x^2+r_y^2}\|\mathbf{r}\|}\begin{pmatrix} -r_xr_z\\ -r_yr_z \\ r_x^2+ r_y^2\end{pmatrix}, \quad \text{and} \\ \quad v &= \frac{1}{\|\mathbf{r}\|}\begin{pmatrix}r_x \\ r_y \\ r_z\end{pmatrix}.
	\end{split}
\end{equation}
Using these three vectors we construct the operators
\begin{equation}\label{eq:mftsymp}
	\begin{split}
	\mu &= \frac{r_y}{\sqrt{r_x^2+r_y^2}\|\mathbf{r}\|} \sigma^x - \frac{r_x}{\sqrt{r_x^2+r_y^2}\|\mathbf{r}\|} \sigma^y, \\
	\nu &= \frac{-r_xr_z}{\sqrt{r_x^2+r_y^2}\|\mathbf{r}\|} \sigma^x - \frac{-r_yr_z}{\sqrt{r_x^2+r_y^2}\|\mathbf{r}\|} \sigma^y + \frac{\sqrt{r_x^2+r_y^2}}{\|\mathbf{r}\|}\sigma^z,\quad \text{and}\\
	\xi &= \frac{r_x}{\|\mathbf{r}\|^2} \sigma^x + \frac{r_y}{\|\mathbf{r}\|^2} \sigma^y +   \frac{r_z}{\|\mathbf{r}\|^2} \sigma^z,
	\end{split}
\end{equation}
which satisfy $\Theta(\mu,\nu) = 1$ and $\Theta(\nu,\xi) = \Theta(\mu,\xi) = 0$. We construct the classical operators via
\begin{equation}
	\begin{split}
		\langle \varphi(f)\rangle &\equiv \tr(\mu\varrho) \int f(x)\,dx = 0,\\
		\langle \pi(f) \rangle &\equiv \tr(\nu\varrho) \int f(x)\,dx = 0, \quad \text{and}\\
		\langle \xi(f) \rangle &\equiv \int f(x)\,dx,
	\end{split}
\end{equation}
and exploit the fact that the expectation value of any product factorises to infer that $\varphi(f) =  \pi(f) = 0$  and $\xi(f) = \int f(x)\,dx$ so that the classical field operators are given by $\varphi(x) = \pi(x) = 0$ and $\xi(x) = 1$. 

The quantum field operators according to the fluctuation field construction are derived from
\begin{equation}
	\begin{split}
		\widetilde{\varphi}^{(a)}(f) &\equiv aZ_{\mu}(a) \sum_{j\in \mathbb{Z}} f(aj) \mu_j,\\
		\widetilde{\pi}^{(a)}(f) &\equiv aZ_{\nu}(a) \sum_{j\in \mathbb{Z}} f(aj) \nu_j, \quad \text{and}\\
		\widetilde{\xi}^{(a)}(f) &\equiv aZ_{\xi}(a) \sum_{j\in \mathbb{Z}} f(aj) \xi_j - Z_{\xi}(a) \int f(x) \,dx\, \mathbb{I}.
	\end{split}
\end{equation}
We note that the choice $Z_A(a) \equiv 1/\sqrt{a}$ for all $A$ is sufficient to ensure that the limit Eq.~(\ref{eq:qctslimit}) always exists. The quantum limit factorises because the operator $\widehat{\xi}(f)$ always commutes with $\widehat{\varphi}(f)$ and $\widehat{\pi}(f)$. Accordingly we obtain a representation of the canonical commutation relations  
\begin{equation}
	[\widehat{\varphi}(f), \widehat{\pi}(g)] = i \int f(x)g(x)\,dx\, \mathbb{I}.
\end{equation}
When we define $\widehat{\psi}(f) = \frac{1}{\sqrt2}(\widehat{\varphi}(f) + i \widehat{\pi}(f))$ for all $f$ we obtain the standard commutation relations
\begin{equation}
	[\widehat{\psi}(f), \widehat{\psi}^\dag(g)] = (f,g)\mathbb{I}
\end{equation}
and learn that the vector $|\Omega\rangle$ is annihilated by $\widehat{\psi}(f)$ for all $f$:
\begin{equation}
	\widehat{\psi}(f)|\Omega\rangle = 0,
\end{equation}
thus we identify the quantum continuum limit as that of a single bosonic field in one spatial dimension. Note that, regardless of the value of $\mathbf{r}$, we obtain the \emph{same} continuum limit description for the quantum fluctuations. The physical intuition for this is relatively simple to explain: the state $\rho_a$ is actually a pure state of the classical continuum limit for all  $\mathbf{r}\not=\mathbf{0}$. (This seemingly paradoxical statement is easily justified: in the classical continuum limit we are focussing on a tiny subset of all observables; when restricting to a reduced set of observables it is the case that mixed states in the bulk of state space can end up on the \emph{boundary} of the effective state space for the reduced set of observables, and hence be effectively pure.) When we study quantum fluctuations we are focussing on an infinitesimal neighbourhood of the classical reference state. The local neighbourhood of any vector $\mathbf{r}\not=\mathbf{0}$ looks the same as any other: it is generically $\mathbb{R}^3$. The only difference is that the symplectic structure on the tangent space is rotated, and the identification of the canonical field operators is correspondingly rotated.

For a quantum spin chain comprised of a chain of $D$-dimensional quantum spins we generically find the quantum continuum limit is given by $\binom{D}{2}$ bosonic quantum fields, $D-1$ classical fields, and the identity field. The explanation for this is relatively easy to understand: first work in the eigenbasis $|j\rangle$ of $\rho$ and construct the single-site hermitian operator basis $\lambda^{(jk)} \equiv \frac12(|j\rangle\langle k|+|k\rangle\langle j|)$, $j,k = 1, 2, \ldots, D$. Next form the $\Theta$ matrix with matrix elements 
\begin{equation}
	[\Theta]_{jk;j'k'} = \delta_{j'k}\delta_{jk'}( r_j - r_k),
\end{equation}
where $r_j$ is the $j$th eigenvalue of $\rho$. Writing $\Theta$ in bra-ket form we find 
\begin{equation}
	\Theta  = \sum_{j\not= k} (r_j-r_k) |jk\rangle\langle kj|.
\end{equation}
In the generic case where all the differences $r_j-r_k \not=0$, $j\not=k$, the kernel of such a matrix is given by all vectors of the form $|jj\rangle$. Thus the rank of such an antisymmetric matrix is generically $D^2-D$ and hence there are $\binom{D}{2}$ canonically conjugate pairs of fluctuation fields $\{(\widehat{\varphi}_j, \widehat{\pi}_j)\,|\, j= 1, 2, \ldots, \binom{D}{2}\}$ obeying the standard canonical commutation relations. 

\subsection{Matrix product states}
In this subsection we study the continuum limit of the matrix product state (MPS) class. We find that our continuum limit construction produces the \emph{continuous matrix product state} class \cite{verstraete:2010a}. An infinite chain of spin-$\frac12$ particles is considered for simplicity; the extension to the general case is sketched afterward. 

Our states $\rho_a$ are given by the sequence
\begin{equation}
	\rho_a = |\psi_a\rangle\langle \psi_a|,
\end{equation}
of translation invariant states, where
\begin{equation}
	|\psi_a\rangle = \sum_{\cdots z_{-1}z_0z_{1} \cdots = 0}^{1}\langle \omega_L| \cdots A^{z_{-1}}_aA^{z_{0}}_a A^{z_{1}}_a\cdots |\omega_R\rangle |\cdots z_{-1}z_0z_{1} \cdots \rangle,
\end{equation}
with the prescription
\begin{equation}
	\begin{split}
		A^0_a &= \mathbb{I} + a Q, \quad \text{and} \\
		A^1_a &= \sqrt{a} R.
	\end{split}
\end{equation}

Define the transfer operator
\begin{equation}
	\mathcal{E}_a(\cdot) = \sum_{z=0}^1 (A^z_a)^\dag (\cdot) A^z_a.
\end{equation}
Expectation values of local observables $M$ are computed according to
\begin{equation}
	\langle M \rangle = \lim_{n\rightarrow \infty}\langle \omega_R| \mathcal{E}^{(n)}_a(\mathcal{F}_a^{M}(\mathcal{E}^{(n)}_a(|\omega_L\rangle \langle \omega_L|)))|\omega_R\rangle / \langle \omega_R| \mathcal{E}^{(2n+1)}_a(|\omega_L\rangle \langle \omega_L|)|\omega_R\rangle,
\end{equation}
where
\begin{equation}
	\mathcal{F}_a^{M}(\cdot) = \sum_{z,z'=0}^1 \langle z|M|z'\rangle  (A^z_a)^\dag (\cdot) A^{z'}_a.
\end{equation}

The state $|\psi_a\rangle$ is invariant under the gauge transformation 
\begin{equation}
A^z \mapsto \sqrt{X} A^{z} \frac{1}{\sqrt{X}}.
\end{equation}
With a judicious choice of $X$ we can always arrange for $\mathcal{E}_a(\mathbb{I}) = \mathbb{I}$ whence the expression for expectation values is simplified to
\begin{equation}
	\langle M \rangle = \tr(\rho_a \mathcal{F}_a^{M}(\mathbb{I})),
\end{equation}
where $\mathcal{E}_a^{\times} (\rho_a) = \rho_a$ and 
\begin{equation}
	\mathcal{E}_a^\times(\cdot) = \sum_{z=0}^1 A^z_a (\cdot) (A^z_a)^\dag.
\end{equation}

The condition $\mathcal{E}_a(\mathbb{I}) = \mathbb{I}$ implies that, to first order in $a$, $Q+Q^\dag + R^\dag R = 0$, so that 
\begin{equation}
	Q = -iK -\frac12 R^\dag R.
\end{equation}
In this case we find that 
\begin{equation}
	\mathcal{E}_a = \mathcal{I} + a \mathcal{L} + O(a^2),
\end{equation}
where
\begin{equation}
	\mathcal{L}(\cdot) = -i[K, \cdot] - \frac12\{R^\dag R, \cdot \} + R (\cdot) R^\dag.
\end{equation}
In this way we see that for $n\lesssim 1/a$:
\begin{equation}
	\mathcal{E}_a^{(n)}(\cdot) \approx e^{an \mathcal{L}} (\cdot).
\end{equation}

We find the classical limit by first computing 
\begin{equation}
	\begin{split}
	\lim_{a\rightarrow 0}\langle M_x \rangle_a &=  \lim_{a\rightarrow 0}\tr(\rho_a \mathcal{F}_a^{M}(\mathbb{I})) =  \lim_{a\rightarrow 0}\sum_{z,z'=0}^1 \langle z|M|z'\rangle  \tr(\rho_a(A^z_a)^\dag A^{z'}_a) \\
	&= \lim_{a\rightarrow 0}  \langle 0|M|0\rangle  + \sqrt{a}[\langle 0|M|1\rangle\tr(\rho_a R) + \langle 1|M|0\rangle \tr(\rho_a R^\dag)] +  a[\langle 1|M|1\rangle  -\langle 0|M|0\rangle]\tr(\rho_a R^\dag R) + O(a^{\frac32}) \\
	&= \langle 0|M|0\rangle.
	\end{split}
\end{equation}
The correlation functions of the classical continuum limit are inferred from
\begin{equation}
	\begin{split}
	\lim_{a\rightarrow 0}\langle M_x M'_y \rangle_a &=  \lim_{a\rightarrow 0}\tr(\rho_a \mathcal{F}_a^{M_x}(\mathcal{E}_a^{\lfloor (y-x)/a \rfloor}((\mathcal{F}_a^{M_y'}(\mathbb{I})))))  \\
	&= \lim_{a\rightarrow 0} \langle 0|M'|0\rangle\tr(\rho_a \mathcal{F}_a^{M_x}(\mathcal{E}_a^{\lfloor (y-x)/a \rfloor}(\mathbb{I}))) + O(\sqrt{a}) \\
	&= \lim_{a\rightarrow 0} \langle 0|M'|0\rangle\tr(\rho_a \mathcal{F}_a^{M_x}(\mathbb{I})) + O(\sqrt{a}) \\
	&= \lim_{a\rightarrow 0} \langle 0|M|0\rangle\langle 0|M'|0\rangle + O(\sqrt{a}) \\
	&= \langle 0|M|0\rangle\langle 0|M'|0\rangle. 
	\end{split}
\end{equation}
All the higher-order correlation functions factorise similarly. Thus the classical continuum limit gives us a pure factorised state. The continuum-limit fields are then simply given by $\phi_x(x) = 0 = \phi_y(x)$ and $\phi_z(x) = 1$.

To construct the quantum continuum limit we need to calculate the $\Theta$ matrix:
\begin{equation}
	\Theta = i\begin{pmatrix} 0 & 1 & 0 \\ -1 & 0 & 0  \\ 0 & 0 & 0  \end{pmatrix}.
\end{equation}
The quantum field operators are then, according to the fluctuation construction, derived from
\begin{equation}
	\begin{split}
		\widetilde{\varphi}^{(a)}(f) &\equiv Z_{\sigma^x}(a) \sum_{j\in \mathbb{Z}} f(aj) \sigma^x_j,\\
		\widetilde{\pi}^{(a)}(f) &\equiv Z_{\sigma^y}(a) \sum_{j\in \mathbb{Z}} f(aj) \sigma^y_j, \quad \text{and}\\
		\widetilde{\xi}^{(a)}(f) &\equiv Z_{\sigma^z}(a) \sum_{j\in \mathbb{Z}} f(aj) \sigma^z_j - Z_{\sigma^z}(a) \int f(x) \,dx\, \mathbb{I}.
	\end{split}
\end{equation}
We note that the choice $Z_A(a) \equiv \sqrt{a}$ for all $A$ is again sufficient to ensure that the limit Eq.~(\ref{eq:qctslimit}) always exists. The quantum limit factorises with respect to $\widehat{\xi}$ because, again, the operator $\widehat{\xi}(f)$ always commutes with $\widehat{\varphi}(f)$ and $\widehat{\pi}(f)$. Accordingly we obtain a representation of the canonical commutation relations  
\begin{equation}
	[\widehat{\varphi}(f), \widehat{\pi}(g)] = i \int f(x)g(x)\,dx\, \mathbb{I}.
\end{equation}
When we define $\widehat{\psi}(f) = \frac{1}{\sqrt2}(\widehat{\varphi}(f) + i \widehat{\pi}(f))$ for all $f$ we again obtain the standard commutation relations
\begin{equation}
	[\widehat{\psi}(f), \widehat{\psi}^\dag(g)] = (f,g)\mathbb{I}.
\end{equation}
However, in contrast to the product-state case,  the vector $|\Omega\rangle$ is not annihilated by $\widehat{\psi}(f)$ for all $f$. Instead, we find that $|\Omega\rangle$ exhibits nontrivial nongaussian correlations. For example, assume that $\supp(f)\cap \supp(g) = \emptyset$ and that the support of $g$ lies to the right of that of $f$.
\begin{equation}
	\begin{split}
	\langle\widehat{\psi}^\dag(f)\widehat{\psi}(g)\rangle &= \lim_{a\rightarrow 0} a\sum_{j,k\in\mathbb{Z}} f(aj)g(ak)\langle \sigma^+_j \sigma^-_k \rangle_a, \\
	&= \lim_{a\rightarrow 0}a\sum_{j,k\in\mathbb{Z}} f(aj)g(ak)\tr(\rho_a \mathcal{F}_a^{ \sigma^+_j}(\mathcal{E}_a^{\lfloor (k-j)/a \rfloor}(\mathcal{F}_a^{ \sigma^-_k}(\mathbb{I})))) \\
	&= \lim_{a\rightarrow 0}a \sqrt{a}\sum_{j,k\in\mathbb{Z}} f(aj)g(ak)\tr(\rho_a \mathcal{F}_a^{ \sigma^+_j}(\mathcal{E}_a^{\lfloor (k-j)/a \rfloor}(R)) + O(a^{\frac{1}{2}})\\
	&= \lim_{a\rightarrow 0}a^2\sum_{j,k\in\mathbb{Z}} f(aj)g(ak)\tr(\rho_a R^\dag (\mathcal{E}_a^{\lfloor (k-j)/a \rfloor}(R)) + O(a^{\frac{1}{2}}) \\
	&= \int f(x) g(y) \tr(\rho_{\text{ss}} R^\dag e^{(y-x)\mathcal{L}}(R))\,dxdy,
	\end{split}
\end{equation}
where $\mathcal{L}^\times(\rho_{\text{ss}}) = 0$. The constraint on the supports of $f$ and $g$ can be lifted upon noticing that
\begin{equation}
	\begin{split}
	\langle\widehat{\psi}^\dag(f)\widehat{\psi}(g)\rangle &= \lim_{a\rightarrow 0} a\sum_{j,k\in\mathbb{Z}} f(aj)g(ak)C_2^{(a)}(j,k), \\
	&= \int f(x) g(y) C_2(x,y)\,dxdy,
	\end{split}
\end{equation}
where 
\begin{equation}
	C_2^{(a)}(\lfloor x/a \rfloor,\lfloor y/a \rfloor) = \langle \sigma^+_{\lfloor x/a \rfloor} \sigma^-_{\lfloor y/a \rfloor} \rangle_a = C_2(x,y) + O(\sqrt{a}),
\end{equation}
with
\begin{equation}
	C_2(x,y) = \begin{cases} \tr(\rho_{\text{ss}} R^\dag e^{(y-x)\mathcal{L}}(R)), \quad &x<y, \\ \tr(\rho_{\text{ss}} R e^{(x-y)\mathcal{L}}(R^\dag)), \quad &x>y. \end{cases}
\end{equation}

Similarly, we obtain the expression
\begin{equation}
	\langle \widehat{\psi}^{\#}(f_1)\widehat{\psi}^{\#}(f_2)\cdots \widehat{\psi}^{\#}(f_n)\rangle = \int f_1(x_1)\cdots f_n(x_n) C_n(x_1^\#, \ldots, x_n^\#)\, dx_1\cdots dx_n,
\end{equation}
where $A^\#$ stands for either $A$ or $A^\dag$ and 
\begin{equation}
	C_n(x_1^\#, \ldots, x_n^\#) \equiv \tr(\rho_{\text{ss}} R^\# e^{(x_1-x_{2})\mathcal{L}}( R^\# e^{(x_{2}-x_{3})\mathcal{L}}(\cdots R^\#)\cdots)),
\end{equation}
for $x_1 \le x_2 \le \cdots \le x_n$. The $n$-point correlation function is extended symmetrically to all other values of the tuple $(x_1, \ldots, x_n)$, i.e., $C_n(x_1^\#, \ldots, x_n^\#) = C_n(x_{\pi^{-1}(1)}^\#, \ldots, x_{\pi^{-1}(n)}^\#)$, where $\pi$ is the permutation that brings the tuple $(x_1, \ldots, x_n)$ into ascending order.

We can now obtain a representation of the quantum field operator $\widehat{\psi}(x)$ by taking for $f$ a sequence of functions tending to a delta function, e.g., $f \equiv \{\chi_\alpha(x)\}$, with 
\begin{equation}
	\chi_\alpha(x) = \begin{cases} 1/\alpha, \quad &x \in [-\alpha, \alpha]\\
	0, \quad &|x| > \alpha.\end{cases}
\end{equation}
In this way we obtain the $n$-point correlation functions
\begin{equation}
	\langle \widehat{\psi}^\#(x_1), \ldots, \widehat{\psi}^\#(x_n)) \equiv \tr(\rho_{\text{ss}} R^\#_1 e^{(x_1-x_{2})\mathcal{L}}( R^\#_{2} e^{(x_{2}-x_{3})\mathcal{L}}(\cdots R_n^\#)\cdots)).
\end{equation}

\subsection{Projected entangled-pair states}
In this subsection we investigate the continuum limit of a nontrivial two-dimensional tensor network, namely, the Projected entangled-pair states (PEPS). Rather than detail the general derivation, we focus on one tractable example \cite{verstraete:2006a} which exemplifies the features of the generic case. This example relies on a mapping to the classical Ising model and allows us exploit classical results for the Ising model. Unfortunately there are still several open mathematical questions concerning the detailed properties of the Ising model, some of which we require for the continuum limit construction. Thus the discussion in this subsection must be regarded as at a physical level of rigour.

Consider an infinite two-dimensional lattice of spin-$\frac12$ particles. Initialise the lattice in the state $|\omega\rangle = \bigotimes_{j\in \mathbb{Z}^2} |+\rangle$, where $|+\rangle = \frac{1}{\sqrt{2}}(|1\rangle + |-1\rangle)$. Let $h_{j,k}$ be the operator acting on sites $j, k\in \mathbb{Z}^2$ as 
\begin{equation}
	h_{j,k}|z_j\rangle_j|z_k\rangle_k = -z_jz_k|z_j\rangle_j|z_k\rangle_k.
\end{equation}
Using these operators we build 
\begin{equation}
	H = \sum_{\langle j, k\rangle} h_{j,k}.
\end{equation}
Now construct the state
\begin{equation}
	|\phi_{\beta}\rangle \equiv \frac{1}{\sqrt{\mathcal{Z}}} e^{-\frac{\beta}{2} H} |\omega\rangle,
\end{equation}
where $\mathcal{Z} = \sum_{z} e^{\beta \sum_{\langle j, k\rangle} z_jz_k}$ is the partition function for the \emph{classical} Ising model at inverse temperature $\beta$. One can check that $\langle \phi_\beta|\phi_\beta\rangle = 1$. It turns out \cite{verstraete:2006a} that $|\phi_\beta\rangle$ is a PEPS with bond dimension $2$. This construction has appeared in the literature in various guises.

All of the $n$-point correlation functions are determined in terms of classical expectation values. Firstly, the $1$-point functions are given by
\begin{equation}
	\langle \phi_\beta|\sigma^z_l|\phi_\beta\rangle = \frac{1}{\mathcal{Z}}\sum_{z} z_le^{\beta \sum_{\langle j, k\rangle} z_jz_k}.
\end{equation}
Similarly,
\begin{equation}
	\langle \phi_\beta|\sigma^x_l|\phi_\beta\rangle = \frac{1}{\mathcal{Z}}\sum_{z}  e^{\beta\sum_{\langle l,m\rangle} z_lz_m} e^{\beta \sum_{\langle j, k\rangle} z_jz_k},
\end{equation}
and
\begin{equation}
	\langle \phi_\beta|\sigma^y_l|\phi_\beta\rangle = -\frac{i}{\mathcal{Z}}\sum_{z} z_l e^{\beta\sum_{\langle l,m\rangle} z_lz_m} e^{\beta \sum_{\langle j, k\rangle} z_jz_k}.
\end{equation}
The calculation of $n$-point functions proceeds similarly.

The correlation length for the class $|\phi_\beta\rangle$ is determined by the correlation length of the classical Ising model. In order to construct the sequence $|\psi_a\rangle$ for the continuum limit construction we need to determine the appropriate lengthscale $a(\beta)$ corresponding to inverse temperature $\beta$. This lengthscale converges to zero as the critical inverse temperature 
\begin{equation}
	\beta_c = \frac{\ln(1+\sqrt{2})}{2}
\end{equation}
is approached. We calculate $a(\beta)$ by first working out the correlation length $\xi$ for the classical Ising model as $\beta_c$ is approached from above/below. The value of $\xi$ is well known:
\begin{equation}
	\xi \sim \frac{1}{1-t},
\end{equation}
with $t = \sinh^4(2\beta)$. Thus we find
\begin{equation}
	a(\beta) = a_0(1-\sinh^4(2\beta)),
\end{equation}
where $a_0$ is an initial lattice spacing.

Now we proceed to the classical continuum limit construction: define for $A\in \mathcal{M}_2(\mathbb{C})$ 
\begin{equation}
	\phi_A^{(a)}(f) = a^2\sum_{j\in \mathbb{Z}^2} f(aj) A_j.
\end{equation}
Taking the limit $\beta \rightarrow \beta_c$ we find the expectation values
\begin{equation}
	\begin{split}
		\phi_A^{(a)}(f) &= \lim_{a\rightarrow0}a^2 \langle \psi_a|A_j|\psi_a\rangle \sum_{j\in \mathbb{Z}^2} f(aj) \\
		&= \left(\lim_{\beta\rightarrow\beta_c} \langle \phi_\beta|A_j|\phi_\beta\rangle\right) \int f(x,y)\,dxdy.
	\end{split}
\end{equation}
We now discuss the quantity $\lim_{\beta\rightarrow\beta_c} \langle \phi_\beta|A_j|\phi_\beta\rangle$. For the case where $A_j = \sigma^z_j$ or $\sigma^y_j$ the spin-flip symmetry $|\psi_\beta\rangle = \bigotimes_{j\in \mathbb{Z}^2} \sigma^x_j|\psi_\beta\rangle$ implies that
\begin{equation}
	\lim_{\beta\rightarrow\beta_c} \langle \phi_\beta|\sigma^z_j|\phi_\beta\rangle = \lim_{\beta\rightarrow\beta_c} \langle \phi_\beta|\sigma^y_j|\phi_\beta\rangle = 0.
\end{equation}
The expectation value of $\sigma_j^x$ is a little more involved. Here we see that the classical observable
\begin{equation}
	S_j^x(z) =  \prod_{\langle j,k\rangle}e^{\beta z_jz_k}
\end{equation} 
which determines $\langle \phi_\beta|\sigma^x_l|\phi_\beta\rangle$ is given, at $\beta\rightarrow \beta_c$ by
\begin{equation}
	S_j^x(z) = \prod_{\langle j,k\rangle}e^{\frac{\ln(1+\sqrt{2})}{2}z_jz_k} = \prod_{\langle j,k\rangle}\sqrt{1+\sqrt{2}}^{z_jz_k}.
\end{equation} 
We note that, crucially, $S_j(z)$ is a \emph{positive} and \emph{bounded} quantity so that the following limit (assumed to exist) is nonzero 
\begin{equation}
	r_x = \lim_{\beta\rightarrow\beta_c} \langle \phi_\beta|\sigma^x_j|\phi_\beta\rangle > 0.
\end{equation}
We now have enough information to calculate the $\Theta$ matrix; we find that $[\Theta]_{jk} = \lim_{\beta\rightarrow\beta_c} \langle \phi_\beta| [\sigma^j, \sigma^k]|\phi_\beta\rangle$ which, in matrix form, is
\begin{equation}
	\Theta = i\begin{pmatrix} 0 & 0 & 0 \\ 0 & 0 & r_x \\ 0 & -r_x & 0\end{pmatrix}.
\end{equation}
The symplectic Gram-Schmidt process immediately delivers $\xi = \sigma^x$, $\mu = -\sqrt{r_x}\sigma^z$, and $\nu = \sqrt{r_x}\sigma^y$.

The quantum continuum limit is determined by the \emph{scaling limit} of the classical Ising model. Thus the result of the continuum limit construction depends on the way the critical point is approached. In this case we obtain a different quantum continuum limit if we approach $T_c$ from above than if we approach it from below. This phenomena is not at all unusual in continuum limits: when approaching a critical point one can obtain many different limits depending on how many relevant operators there are.

The quantum field operators are then, according to the fluctuation construction, derived from
\begin{equation}
	\begin{split}
		\widetilde{\varphi}^{(a)}(f) &\equiv Z_{\mu}(a) \sum_{j\in \mathbb{Z}^2} f(aj) \mu_j,\\
		\widetilde{\pi}^{(a)}(f) &\equiv Z_{\nu}(a) \sum_{j\in \mathbb{Z}^2} f(aj) \nu_j, \quad \text{and}\\
		\widetilde{\xi}^{(a)}(f) &\equiv Z_{\xi}(a) \sum_{j\in \mathbb{Z}^2} f(aj) \xi_j - r_xZ_{\xi}(a) \int f(x,y) \,dxdy\, \mathbb{I}.
	\end{split}
\end{equation}
The choice of $Z_{\mu}(a)$ is now dictated by two constraints: (i) the limits Eq.~(\ref{eq:qctslimit}) should all exist; and (ii) the pair $(\widetilde{\varphi}^{(a)}(f),\widetilde{\pi}^{(a)}(f) )$ should, in the limit $a\rightarrow 0$, obey the canonical commutation relations. There is a tension between these two requirements and, in the case where there is an incompatibility between these two requirements we are content to relax the first requirement to apply only to the $n$-point functions of the field strength operator $\widetilde{\varphi}^{(a)}(f)$ and fix $\widetilde{\pi}^{(a)}(f)$ by the second requirement. 

Let's now investigate what scaling $Z_{\mu}(a)$ will ensure the existence of the $n$-point functions. Actually we won't discuss this problem in full generality and only study the $2$-point function case as the $n$-point case is no more complicated. Consider
\begin{equation}
	\begin{split}
	\lim_{a\rightarrow 0} \tr\left(\rho_a \widetilde{\varphi}^{(a)}(f)\widetilde{\varphi}^{(a)}(g)\right) &= \lim_{a\rightarrow 0}r_x^2Z_{\mu}^2(a) \sum_{j,k\in \mathbb{Z}^2} f(aj)g(ak) \langle \psi_a|\sigma^z_j\sigma^z_k|\psi_a\rangle \\
	&= \lim_{a\rightarrow 0}r_x^2Z_{\mu}^2(a) \sum_{j,k\in \mathbb{Z}^2} f(aj)g(ak) \langle\psi_a|\sigma^z_0\sigma^z_{k-j}|\psi_a\rangle.
	\end{split}
\end{equation}
Now, it is known \cite{mccoy:1977a} that, for all $(x,y)\in \mathbb{R}^2$, the following (rotation invariant) limit exists
\begin{equation}
	G_{\pm}(x,y) = \lim_{a\rightarrow 0} a^{-\frac{1}{4}} \langle \sigma^z_{(0,0)}\sigma^z_{(\lfloor x/a\rfloor,\lfloor y/a\rfloor)}\rangle.
\end{equation}
Exploiting this information we see that, if we choose $Z_\mu(a) = a^{\frac{15}{8}}$, we obtain a well-defined limit
\begin{equation}
	\begin{split}
	\lim_{a\rightarrow 0} \tr\left(\rho_a \widetilde{\varphi}^{(a)}(f)\widetilde{\varphi}^{(a)}(g)\right) &= \lim_{a\rightarrow 0}r_x^2a^{\frac{15}{4}} \sum_{j,k\in \mathbb{Z}^2} f(aj)g(ak) \langle \psi_a|\sigma^z_j\sigma^z_k|\psi_a\rangle \\
	&= \lim_{a\rightarrow 0}r_x^2a^{4} \sum_{j,k\in \mathbb{Z}^2} f(aj)g(ak) a^{-\frac14}\langle \psi_a|\sigma^z_0\sigma^z_{k-j}|\psi_a\rangle \\
	&= \lim_{a\rightarrow 0}r_x^2a^{4} \sum_{j,k\in \mathbb{Z}^2} f(aj)g(ak) G_{\pm}(a(k-j)) \\
	&= r_x^2 \int f(x)g(y)G_{\pm}(y-x) \, dxdy.
	\end{split}
\end{equation}
It is known \cite{mccoy:1977a} that with this choice of $Z_\mu(a)$ all higher-order correlation functions involving $\widetilde{\varphi}^{(a)}$ exist.
In order to maintain the canonical commutation relations we are then forced to choose  
\begin{equation}
	Z_\mu(a)Z_\nu(a) = a^2, \quad \text{i.e.} \quad Z_{\nu}(a) = a^{\frac18}.
\end{equation}
Thus we obtain \emph{two} possible quantum continuum limits $|\Omega_{\pm}\rangle$ for the Ising PEPS, with
\begin{equation}
	\langle \Omega_{\pm}|\widehat{\varphi}(0)\widehat{\varphi}(x)|\Omega_{\pm}\rangle = G_{\pm}(x).
\end{equation}
Because the functions $Z_{\mu}(a)$ and $Z_{\nu}(a)$ differ there is an arbitrariness in the construction of annihilation and creation field operators. 

\subsection{Tree tensor network states}
Here we apply the continuum limit construction to a sequences of states derived from \emph{tree tensor networks} (TTN), a class of states arising from the application of the Kadanoff block-spin renormalisation group. Again, for simplicity, we focus on a one-dimensional chain of $d$-dimensional quantum spins; there is no difficulty in generalising these results to higher dimensions. Remarkably, the construction described here also works essentially without modification in the more general case of the multiscale entanglement renormalisation ansatz (MERA) \cite{vidal:2007a}. 

Our TTN is defined as follows. First fix an isometry $V:\mathbb{C}^d\otimes \mathbb{C}^d \rightarrow \mathbb{C}^d$ and build the larger isometry
\begin{equation}
	\mathcal{V} \equiv \bigotimes_{k\in 2\mathbb{Z}} V
\end{equation}
on the spin chain.
The matrix elements of the action of $V$ on operators are given by
\begin{equation}
	V\mu^{\alpha}\otimes\mu^{\beta} V^\dag = \sum_{\gamma = 0}^3 {v^{\alpha\beta}}_{\gamma} \mu^\gamma,
\end{equation}
where $\mu^\alpha$ is a complete orthonormal basis of operators with respect to the Hilbert-Schmidt inner product: $(\mu^\alpha, \mu^\beta) \equiv \frac1d\tr({\mu^\alpha}^\dag \mu^\beta) = \delta^{\alpha\beta}$. We typically assume that $\mu^0 \equiv \mathbb{I}$.
Our subsequent discussion is simplified somewhat when we make the additional assumption that $V$ is \emph{symmetric}, i.e., 
\begin{equation}
	V\textsc{swap} = V,
\end{equation}
where
\begin{equation}
	\textsc{swap} = \sum_{jk} |kj\rangle\langle jk|.
\end{equation}

Using $\mathcal{V}$ we construct the CP map
\begin{equation}
	\mathcal{E}(M) \equiv \frac12 \mathcal{V}M\mathcal{V}^\dag + \frac12 \mathcal{V}\mathcal{T}M\mathcal{T}^\dag\mathcal{V}^\dag,
\end{equation}
where $\mathcal{T}$ is the ``shift right'' automorphism:
\begin{equation}
	\mathcal{T}|\cdots \rangle |a\rangle_{-1} |b\rangle_{0} |c\rangle_{1} |\cdots \rangle = |\cdots \rangle |a\rangle_{0} |b\rangle_{1} |c\rangle_{2} |\cdots \rangle.
\end{equation}
The channel $\mathcal{E}$ is then used to prepare the sequence of states
\begin{equation}
	\rho_{\ell} \equiv (\mathcal{E}^\times)^{(\ell)}(\rho_0),
\end{equation}
where $\mathcal{E}^\times(\sigma)  \equiv \frac12 \mathcal{V}^\dag\sigma\mathcal{V} + \frac12 \mathcal{T}^\dag\mathcal{V}^\dag\sigma\mathcal{V}\mathcal{T}$ and 
\begin{equation}
	\rho_0 \equiv \bigotimes_{j\in\mathbb{Z}} |\omega_0\rangle\langle\omega_0|.
\end{equation}

A special role is played by the \emph{primary operators} which are defined by
\begin{equation}
	\mathcal{E}(\phi_{j,j+1}) = \begin{cases} \lambda \phi_{j/2, j/2+1}, \quad \text{$j$ even}, \\ \lambda \phi_{(j-1)/2,j/2} \quad \text{$j$ odd},\end{cases} 
\end{equation}
written, compactly, 
\begin{equation}
	\mathcal{E}(\phi_{j,j+1}) = \lambda \phi_{\lfloor j/2 \rfloor, \lfloor j/2 \rfloor+1}.
\end{equation}
The operators $\phi$ can be found by solving the eigenvalue equation
\begin{equation}\label{eq:ttntx}
	\mathcal{A}(\phi) \equiv \frac{1}{2}V \phi V^\dag\otimes \mathbb{I} + \frac12V\otimes V(\mathbb{I}\otimes\phi\otimes \mathbb{I}) V^\dag\otimes V^\dag =    \lambda\phi,
\end{equation} 
where the CP map $\mathcal{A}$ is called the \emph{ascending operator}. In the symmetric case a subclass of the primary operators can be constructed in terms of $\psi$, given by
\begin{equation}
	\mathcal{A}_L(\psi) \equiv V(\psi\otimes \mathbb{I})V^\dag = \lambda \psi,
\end{equation}
where the CP map $\mathcal{A}_L$ is called the \emph{left ascending operator}. 
Given such a $\psi$, we can build a primary operator $\phi$ simply via
\begin{equation}
	\phi \equiv \psi\otimes \mathbb{I}.
\end{equation}
We call these operators the \emph{simple primary operators}.

The matrix elements of the primary operator eigenvector equation are given by
\begin{equation}
	\frac12\sum_{\alpha,\beta,\gamma} \phi_{\alpha,\beta} {v^{\alpha,\beta}}_{\gamma} \delta_{0,\gamma'} + \frac12 \sum_{\alpha,\beta} \phi_{\alpha,\beta} {v^{0,\alpha}}_{\gamma} {v^{\beta,0}}_{\gamma'} = \lambda\phi_{\gamma,\gamma'},
\end{equation}
which is compactly summarised in vector form by $\mathbb{A}|\phi\rangle = \lambda |\phi\rangle$, where $|\phi\rangle = \sum_{\alpha,\beta} \phi_{\alpha,\beta}|\alpha\rangle|\beta\rangle$, and
\begin{equation}
	\mathbb{A} = \sum_{\alpha,\beta,\alpha',\beta'} \left(\frac{1}{2} {v^{\alpha,\beta}}_{\alpha'} \delta_{0,\beta'} + \frac12{v^{0,\alpha}}_{\alpha'} {v^{\beta,0}}_{\beta'}\right)|\alpha'\beta'\rangle\langle\alpha\beta|.
\end{equation}
Generically there is a unique eigenvector with eigenvalue $1$, namely, $\phi = \mathbb{I}\otimes \mathbb{I}$ and the remaining eigenvectors all have eigenvalue $|\lambda| < 1$. We henceforth assume we are in this generic case. Because $\mathbb{A}$ is not generically a normal matrix we must exploit the Jordan canonical form to diagonalise it. To simplify our discussion somewhat we assume that the matrix is diagonalisable (it turns out that all nontrivial Jordan blocks are averaged away by the continuum limit procedure anyway). Thus we have the eigenvalue decomposition
\begin{equation}
	\mathbb{A} = \sum_{\gamma} \lambda_\gamma |\phi_\gamma\rangle\langle \xi_\gamma|.
\end{equation}

Expectation values for local observables with respect to the sequence $\rho_{\ell}$ can now be found as follows. Suppose we are interested in some observable $A$ which, for concreteness, acts only on sites $j$ and $j+1$. Then, exploiting the \emph{completeness} (but not \emph{orthonormality}!) of the eigenbasis $\phi^\alpha_{j,j+1}$ of Eq.~(\ref{eq:ttntx}), we deduce that there are $a_\alpha$ such that
\begin{equation}
	A_{j,j+1} = \sum_{\alpha=0}^{d^4-1} a_\alpha \phi^\alpha_{j,j+1}.
\end{equation}
The expectation value of $A_{j,j+1}$ then follows from
\begin{equation}
	\begin{split}
		\langle A_{j,j+1} \rangle_{\ell} &= \tr(\rho_{\ell} A_{j,j+1}) = \tr(\rho_0 \mathcal{E}^{(\ell)}(A_{j,j+1})) \\
		&= \sum_{\alpha=0}^{d^4-1} a_\alpha \lambda_{\alpha}^\ell \tr(\rho_0\phi^\alpha_{\lfloor j/2^\ell \rfloor, \lfloor j/2^\ell \rfloor+1}).
	\end{split}
\end{equation}
In the generic case where $\lambda_0$ is the only eigenvalue with $|\lambda_0| = 1$ we deduce that 
\begin{equation}
	\lim_{\ell\rightarrow \infty}\langle A_{j,j+1} \rangle_{\ell}  =  a_0\tr(\rho_0\phi^0_{0,1}).
\end{equation}
Since $\phi^\alpha$ is a complete basis we shift our attention to this set of operators rather than the operator basis $\mu^\alpha\otimes \mu^\beta$.

The $n$-point correlation functions for $\rho_\ell$ may be computed similarly. Suppose that $j_1 < j_2 < \cdots < j_n$ are the locations of the $n$ operators involved in a $n$-point correlator. Note first that if $|j_{k+1}-j_{k}| > 3$ for all $k = 1, 2, \ldots, n-1$, then
\begin{equation}
	\mathcal{E}\left(\prod_{k=1}^n \phi_{j_k,j_{k}+1}^{\alpha_k}\right) = \prod_{k=1}^n \lambda_{\alpha_k} \phi_{\lfloor j_k/2 \rfloor,\lfloor j_k/2 \rfloor+1}^{\alpha_k}.
\end{equation}
If $|j_{k+1}-j_{k}| \ge 2^{\ell+1}$ for all $k = 1, 2, \ldots, n-1$ then the correlator is exactly a product:
\begin{equation}
	\left\langle \prod_{k=1}^n \phi_{j_k,j_{k}+1}^{\alpha_k} \right\rangle_{\ell}  = \prod_{k=1}^n \lambda_{\alpha_k}^{\ell} \tr(\rho_0  \phi^{\alpha_k}).
\end{equation}
The correlation length of the state $\rho_\ell$ is $\xi_\ell = 2^{\ell+1}$.
A key role in the calculation of the $n$-point functions is played by the \emph{fusion maps} $\mathcal{F}_e$ and $\mathcal{F}_o$, defined by 
\begin{equation}
	\mathcal{F}_e(\phi^\alpha\otimes \phi^{\beta}) =  \mathcal{E}\left(\mathcal{E}\left(\phi^\alpha_{0,1}\otimes \phi^{\beta}_{2,3}\right)\right)
\end{equation}
and
\begin{equation}
	\mathcal{F}_o(\phi^\alpha\otimes \phi^{\beta}) =  \mathcal{E}\left(\mathcal{E}\left(\phi^\alpha_{0,1}\otimes \phi^{\beta}_{3,4}\right)\right).
\end{equation}
The action of the maps $\mathcal{F}_e$ and $\mathcal{F}_o$ may be represented in terms of the primary operators as
\begin{equation}
	{F_e^{\alpha\beta}}_\gamma \equiv \tr\left ((\xi_{0,1}^\gamma)^\dag\mathcal{E}\left(\mathcal{E}\left(\phi^\alpha_{0,1}\otimes \phi^{\beta}_{2,3}\right)\right)\right)
\end{equation}
and
\begin{equation}
	{F_o^{\alpha\beta}}_\gamma \equiv \tr\left ((\xi_{0,1}^\gamma)^\dag\mathcal{E}\left(\mathcal{E}\left(\phi^\alpha_{0,1}\otimes \phi^{\beta}_{3,4}\right)\right)\right).
\end{equation}
Using the fusion maps we can calculate the lattice $2$-point function
\begin{equation}
	\tr(\rho_{\ell}\phi_{j,j+1}^{\alpha}\phi_{k,k+1}^{\beta}).
\end{equation}
Supposing that $j<k$ and $|k-j| < 2^\ell$, and using translation invariance, we reduce our problem to calculating
\begin{equation}
	\tr(\rho_{\ell}\phi_{j,j+1}^{\alpha}\phi_{k,k+1}^{\beta}) = \tr(\rho_{\ell}\phi_{0,1}^{\alpha}\phi_{k-j,k-j+1}^{\beta}).
\end{equation}
Let $m\ge 0$ be such that 
\begin{equation}
	k-j = 2^m + 2^{m-1}r_1 + r_0,
\end{equation}
where $0 \le r_0 < 2^{m-2}$ and $r_1 = \lfloor \frac{r}{2^{m-1}}\rfloor \in  \{0,1\}$.
Thus
\begin{equation}
	\tr(\rho_{\ell}\phi_{0,1}^{\alpha}\phi_{k-j,k-j+1}^{\beta}) = (\lambda_\alpha\lambda_\beta)^{m-1}\tr(\rho_{\ell-m+1} \phi_{0,1}^\alpha \phi_{2 + r_1, 3+r_1}^\beta).
\end{equation}
There are two cases:
\begin{equation}
	\tr(\rho_{\ell}\phi_{0,1}^{\alpha}\phi_{k-j,k-j+1}^{\beta}) = \begin{cases} (\lambda_\alpha\lambda_\beta)^{m-1}(\lambda_\gamma)^{\ell-m-1}{F_e^{\alpha\beta}}_\gamma\tr(\rho_{0} \phi_{0,1}^\gamma), &\quad \text{$r_1=0$,} \\ 
	(\lambda_\alpha\lambda_\beta)^{m-1}(\lambda_\gamma)^{\ell-m-1}{F_o^{\alpha\beta}}_\gamma\tr(\rho_{0} \phi_{0,1}^\gamma), &\quad \text{$r_1=1$.}\end{cases}
\end{equation}

Define the following \emph{scaling} function. Write for $x\in \mathbb{R}$
\begin{equation}
	\lfloor x/a \rfloor = 2^m + 2^{m-1}r_1 + r_0.
\end{equation} 
Note the dependence of the values of $m$ and $r_1$ on $x$: 
\begin{equation}
	m(x) = \lfloor \log_2(x/a) \rfloor
\end{equation}
and
\begin{equation}
	r_1(x) = \lfloor \lfloor x/a \rfloor 2^{-m(x/a)+1} - 2 \rfloor.
\end{equation}
Define
\begin{equation}
	C_{\alpha\beta}^{(a)}(x) \equiv \tr(\rho_{\ell}\phi_{0,1}^{\alpha}\phi_{\lfloor x/a \rfloor,\lfloor x/a \rfloor+1}^{\beta}) =  (\lambda_\alpha\lambda_\beta)^{m-1}(\lambda_\gamma)^{\ell-m-1}{F_a^{\alpha\beta}}_\gamma(x)\tr(\rho_{0} \phi_{0,1}^\gamma), 
	\end{equation}
where
\begin{equation}
	{F_a^{\alpha\beta}}_\gamma(x) =  \begin{cases} {F_e^{\alpha\beta}}_\gamma, &\quad \text{$\lfloor \lfloor x/a \rfloor 2^{-\lfloor \log_2(x/a) \rfloor+1} - 2 \rfloor=0$,} \\ 
	{F_o^{\alpha\beta}}_\gamma, &\quad \text{$\lfloor \lfloor x/a \rfloor 2^{-\lfloor \log_2(x/a) \rfloor+1} - 2 \rfloor=1$.}\end{cases}
\end{equation}
The function ${F_a^{\alpha\beta}}_\gamma(x)$ enjoys a discrete scaling invariance:
\begin{equation}
	{F_a^{\alpha\beta}}_\gamma(2 x) = {F_a^{\alpha\beta}}_\gamma(x).
\end{equation}

We are interested in the scaling, for fixed $x$, of $C_{\alpha\beta}^{(a)}(x)$ as $a\rightarrow 0$. Write $\mu = \lambda_\alpha\lambda_\beta$ and notice that
\begin{equation}	
	\mu^{m} = e^{m\log(\mu)}	= e^{\lfloor \log_2(x/a)\rfloor \log(\mu)} = e^{\log_2(x/a) \log(\mu)}\mu^{\lfloor \log_2(x/a)\rfloor-\log_2(x/a) } = M_a(x) (x/a)^{\log_2(\mu)},
\end{equation}
where the function $M_a(x)$ oscillates between $1$ and $1/\mu$. This function also has a discrete scale invariance:
\begin{equation}
	M_a(2x) = M_a(x).
\end{equation}
Thus for fixed $|x| \le 1$ we see that
\begin{equation}
|\mu^{m}| \underset{a\rightarrow 0}{\sim} a^{-\log_2(\mu)}.
\end{equation}
Similarly,
\begin{equation}	
	\lambda_\gamma^{\ell-m} = e^{(\ell-m)\log(\lambda_\gamma)} = e^{(\lfloor \log_2(1/a)\rfloor-\lfloor \log_2(x/a)\rfloor) \log(\lambda_\gamma)} = e^{- \log_2(x) \log(\lambda_\gamma)}N_a(x) = N_a(x)x^{-\log_2(\lambda_\gamma)},
\end{equation}
where
\begin{equation}
	N_a(x) = \lambda_\gamma^{(\lfloor \log_2(1/a)\rfloor-  \log_2(1/a))-(\lfloor \log_2(x/a)\rfloor -\log_2(x/a))}.
\end{equation}
Note that $N_a(x)$ oscillates between $\lambda_\gamma$ and $\frac{1}{\lambda_\gamma}$ and also shares the discrete scale invariance property:
\begin{equation}
	 N_a(2x) = N_a(x). 
\end{equation}
For fixed $|x| \le1$ it follows that
\begin{equation}
	|\lambda_\gamma^{\ell-m}| \underset{a\rightarrow 0}{\sim} O(1).
\end{equation}
Putting this together we learn that, for fixed $|x| \le 1 $,
\begin{equation}
	\tr(\rho_{\ell}\phi_{0,1}^{\alpha}\phi_{\lfloor x/a \rfloor,\lfloor x/a \rfloor+1}^{\beta}) =  \frac{a^{-\log_2(\lambda_\alpha) -\log_2(\lambda_\beta)}}{\lambda_\alpha\lambda_\beta\lambda_\gamma} x^{\log_2(\lambda_\alpha)+\log_2(\lambda_\beta)-\log_2(\lambda_\gamma)}{G_a^{\alpha\beta}}_\gamma(x)\tr(\rho_{0} \phi_{0,1}^\gamma),
\end{equation}
where
\begin{equation}
	{G_a^{\alpha\beta}}_\gamma(x) = M_a(x)N_a(x){F_a^{\alpha\beta}}_\gamma(x).
\end{equation}
The correlation function scales as
\begin{equation}
	|C_{\alpha\beta}^{(a)}(x)| \underset{a\rightarrow 0}{\sim} a^{-\log_2(\lambda_\alpha)}a^{-\log_2(\lambda_\beta)}.
\end{equation}
Notice that along the subsequence $a_j = 2^{-j}$
\begin{equation}
	{G_{a_j}^{\alpha\beta}}_\gamma(x) = \left(\frac{\lambda_\alpha\lambda_\beta}{\lambda_\gamma}\right)^{\lfloor \log_2(x)\rfloor-\log_2(x) }{F^{\alpha\beta}}_\gamma(x)
\end{equation}
where
\begin{equation}
	{F^{\alpha\beta}}_\gamma(x) =  \begin{cases} {F_e^{\alpha\beta}}_\gamma, &\quad \text{$\lfloor \lfloor x \rfloor 2^{-\lfloor \log_2(x) \rfloor+1} - 2 \rfloor=0$,} \\ 
	{F_o^{\alpha\beta}}_\gamma, &\quad \text{$\lfloor \lfloor x \rfloor 2^{-\lfloor \log_2(x) \rfloor+1} - 2 \rfloor=1$.}\end{cases}
\end{equation}
Note that along the subsequence $a_j= 2^{-j}$ the quantity ${G_{a}^{\alpha\beta}}_\gamma(x)$ doesn't depend on $a$.

Hence we obtain the limit
\begin{equation}
	\lim_{a\rightarrow 0} a^{\log_2(\lambda_\alpha)+ \log_2(\lambda_\beta)}\tr(\rho_{\ell}\phi_{0,1}^{\alpha}\phi_{\lfloor x/a \rfloor,\lfloor x/a \rfloor+1}^{\beta}) = \frac{1}{\lambda_\alpha\lambda_\beta\lambda_\gamma} x^{\log_2(\lambda_\alpha)+\log_2(\lambda_\beta)-\log_2(\lambda_\gamma)}{G^{\alpha\beta}}_\gamma(x)\tr(\rho_{0} \phi_{0,1}^\gamma).
\end{equation}

\subsubsection{The classical continuum limit}
The structure of a tree tensor network is covariant with respect to a discrete scaling of the position coordinate by a factor of a $2$: after such a scaling we have again a valid TTN for the lattice. This leads to certain peculiarities for the $n$-point functions in the limit where the lattice spacing goes to $0$: the correlation functions do not tend to a continuous limit but oscillate around some value. This is due to the functions $M_a(x)$ and $N_a(x)$. To overcome this we simply restrict to a convergent subsequence, specifically, $a_j\equiv 2^{-j}$, $j=0,1,2,\ldots$. Thus, from now on, any limit of the form
\begin{equation}
	\lim_{a\rightarrow 0}
\end{equation}
is to be interpreted as the limit 
\begin{equation}
	\lim_{j\rightarrow \infty}
\end{equation}
along the subsequence $a \equiv a_j = 2^{-j}$.

(One could imagine instead using a modified prescription for the sequence $\rho_{\ell(a)}$, e.g.,
\begin{equation}
	\rho_{a} \equiv \frac{1}{\log(2)}\int_{a}^{a/2} \rho_{\lfloor \log_2(1/s)\rfloor}\, \frac{ds}{s}.
\end{equation}
This definition is equivalent to smearing out the field operators over a lattice with lattice spacing chosen at random from the interval $[a/2, a]$. This does lead to a limit but, unfortunately, the resulting expressions for the correlation functions are somewhat awkward.)

The classical continuum limit is now straightforward to describe: let $A_{j,j+1} = \sum_{\alpha=0}^{d^4-1} a_\alpha \phi^\alpha_{j,j+1}.$
\begin{equation}
	\begin{split}
	\langle \phi_\alpha(f) \rangle  &\equiv 2\lim_{a\rightarrow 0}\langle\phi_{\alpha}^{(a)}(f) \rangle =  \lim_{a\rightarrow 0} a\sum_{j\in\mathbb{Z}} f(aj) \tr(\rho_{\lfloor \log_2(1/a)\rfloor}\phi_{j,j+1}^\alpha) \\
	&= \lim_{a\rightarrow 0} a\sum_{j\in\mathbb{Z}} \lambda_\alpha^{\lfloor \log_2(1/a)\rfloor}f(aj) \tr(\rho_0\phi_{j,j+1}^\alpha)  \\
	&= \lim_{a\rightarrow 0}  a\sum_{j\in\mathbb{Z}}f(aj)  \tr(\rho_0\phi_{0,1}^0) \delta_{\alpha,0}\\
	&= \delta_{\alpha,0}\tr(\rho_0\phi_{0,1}^0) \int_{-\infty}^{\infty} f(x)\,dx.  \\
	\end{split}
\end{equation}
In the generic case all the classical $n$-point functions factorise.

\subsubsection{The quantum continuum limit}
The correlation functions for the quantum continuum limit are essentially determined by the $2$-point functions. 

Firstly, the theta matrix is constructed for the simple primary operators via
\begin{equation}
	[\Theta]_{\alpha,\beta} = \lim_{\ell\rightarrow \infty} \langle\omega_0| \mathcal{A}^{(\ell)}([\psi^\alpha,\psi^\beta])|\omega_0\rangle,
\end{equation} 
where the $\{\psi^\alpha\}$ is the set of all the simple primary operators. If we want to include all the other primary operators in the continuum limit the expression is a little more complicated:
\begin{equation}
	[\Theta]_{\alpha,\beta} = \frac{1}{3}\lim_{\ell\rightarrow \infty} \tr(\rho_0 \mathcal{E}^{(\ell)}([\phi^\alpha_{0,1},\phi^\beta_{-1,0}+\phi^\beta_{0,1}+\phi^\beta_{1,2}])).
\end{equation}
Applying the symplectic Gram-Schmidt process produces the canonical discretised field operators:
\begin{equation}
	\begin{split}
		\widetilde{\varphi}^{(a)}_\gamma(f) &\equiv 2Z_{\mu^{\gamma}}(a) \sum_{j\in \mathbb{Z}} (\mu_j^{\gamma} - \langle   \mu_j^{\gamma}\rangle \mathbb{I}_j)\int_{a/2}^a s f(sj) \,ds,\\
		\widetilde{\pi}^{(a)}_{\gamma}(f) &\equiv 2Z_{\nu^{\gamma}}(a) \sum_{j\in \mathbb{Z}}  (\nu_j^{\gamma} - \langle   \nu_j^{\gamma}\rangle \mathbb{I}_j)\int_{a/2}^a s f(sj)\,ds, \quad \text{and}\\
		\widetilde{\xi}^{(a)}_{\gamma}(f) &\equiv 2Z_{\xi^{\gamma}}(a) \sum_{j\in \mathbb{Z}}  (\xi_j^{\gamma} - \langle \xi_j^{\gamma}\rangle \mathbb{I}_j)\int_{a/2}^a s f(sj) \,ds.
	\end{split}
\end{equation}
In general the canonical discretised field operators will be linear combinations of the primary operators associated with different eigenvalues of the ascending operator. 

To simplify the discussion of the $2$-point function we simply assume that the canonical field operators $\widetilde{\varphi}^{(a)}(f)$ are directly associated with simple primary operators, i.e., $\mu_j^{\gamma}$ is a simple primary operator. (This turns out to be the case in many examples). The general case is only slightly more involved.

Let $Z_{\psi^{\alpha}}(a) = a^{\log_2(\lambda_\alpha)+1}$. Then
\begin{equation}
	\begin{split}
	\langle \widetilde{\varphi}^{(a)}_{\alpha}(f)\widetilde{\varphi}^{(a)}_{\beta}(g)\rangle &= 4a^2Z_{\psi^{\alpha}}(a)Z_{\psi^{\beta}}(a) \sum_{j,k}  f(aj)g(ak)\langle \psi_j^{\alpha}\psi_k^{\beta} \rangle \\
	&=  4a^{\log_2(\lambda_\alpha)+\log_2(\lambda_\beta)+2}\sum_{j,k} f(aj)g(ak)C_{\alpha\beta}^{(a)}(a(k-j))   \\
	&\underset{a\rightarrow 0}{\longrightarrow}   \frac{4\tr(\rho_{0} \phi_{0,1}^\gamma)}{\lambda_\alpha\lambda_\beta\lambda_\gamma}\int_{-\infty}^\infty \int_{-\infty}^\infty f(x)g(y) (y-x)^{\log_2(\lambda_\alpha)+\log_2(\lambda_\beta)-\log_2(\lambda_\gamma)}{G^{\alpha\beta}}_\gamma(y-x) \,dxdy.
	\end{split}
\end{equation}

\subsubsection{A simple example}
Here we apply the quantum continuum limit construction to the following simple example:
\begin{equation}
	V = |0\rangle \langle \psi_p| + |1\rangle \langle \phi^+|,
\end{equation}
where
\begin{equation}
	|\psi_p\rangle = \sqrt{p}|00\rangle + \sqrt{1-p}|11\rangle, \quad\text{and}\quad |\phi^+\rangle = \frac{1}{\sqrt{2}}(|01\rangle + |10\rangle).
\end{equation}
The intuition behind this choice is as follows. In this state there is only one type of excitation (labelled ``1''). The first vector in the isometry describes the nontrivial fusion of two excitations: two excitations can fuse to produce the vacuum and the second vector describes how a single excitation behaves under changes of scale.

The sequence of states 
\begin{equation}
	\rho_{\ell} \equiv (\mathcal{E}^\times)^{(\ell)}(\rho_0),
\end{equation}
with $|\omega_0\rangle = |0\rangle$
given by this prescription can alternatively be described as follows: one can think of the action of term $|\psi_p\rangle\langle 0|$ in $V^\dag$ as nucleating a pair of magnons with probability $(1-p)/2$ and the term $|\phi^+\rangle\langle 1|$ as a scale transformation of the magnon. Thus we have magnons being created at the rate $(1-p)/2$ which then spreading throughout the chain as we zoom in. 

The simple scaling operators are given by
\begin{equation}
	\psi^0 = \mathbb{I}, \quad \psi^{1} = \tfrac{1/2-p}{3/2-p}\mathbb{I} + \sigma^z, \quad \psi^2 = \sigma^x, \quad\text{and}\quad \psi^{3} = \sigma^y.
\end{equation}
The corresponding eigenvalues are 
\begin{equation}
	\lambda_0 = 1, \quad \lambda_{1} = \tfrac{p-(1-p)}{2}, \quad \lambda_2 = \tfrac{\sqrt{p}+\sqrt{1-p}}{\sqrt{2}}, \quad\text{and}\quad \lambda_{3} = \tfrac{\sqrt{p}-\sqrt{1-p}}{\sqrt{2}}.
\end{equation}

The right eigenvector with eigenvalue $1$ of the ascending operator is 
\begin{equation}
	\rho_0 = \frac{\mathbb{I} + p'\sigma^z}{2}.
\end{equation}
where
\begin{equation}
	p' = \frac{p-1/2}{3/2-p}.
\end{equation}

As usual the classical fields are then 
\begin{equation}
	\phi_\alpha^{(a)}(f) = a\sum_{j\in \mathbb{Z}} f(aj) \psi^{\alpha}_j.
\end{equation}
with expectation values in the limit $a\rightarrow 0$: 
\begin{equation}
	\langle\phi_\alpha(f)\rangle = \lim_{a\rightarrow 0}a\sum_{j\in \mathbb{Z}} f(aj) \langle \psi^{\alpha}_j\rangle_{\ell = \lfloor 1/a\rfloor} = \delta_{0,\alpha} \int f(x)\, dx.
\end{equation}
The theta matrix for the simple scaling operators is given by
\begin{equation}	
	[\Theta]_{\alpha\beta} = \tr(\rho_0 [\psi^\alpha,\psi^\beta]) \equiv 4i\begin{pmatrix} 0 & 0 & 0 & 0\\ 0 & 0 & 0 & 0\\ 0 & 0 & 0 & p'\\ 0 & 0 & -p' & 0\end{pmatrix}.
\end{equation}
Hence we introduce the quantum fluctuation fields 
\begin{equation}
	\begin{split}
		\widetilde{\varphi}^{(a)}(f) &= Z_{\varphi}(a)\sum_{j\in \mathbb{Z}} f(aj) \sigma^x_j, \quad \text{and} \\
		\widetilde{\pi}^{(a)}(f) &= Z_{\pi}(a)\sum_{j\in \mathbb{Z}} f(aj) \sigma^y_j.
	\end{split}
\end{equation}
The commutation relations are determined by
\begin{equation}
	[\widetilde{\varphi}^{(a)}(f), \widetilde{\pi}^{(a)}(g)] = 2iZ_{\varphi}(a)Z_{\pi}(a)\sum_{j\in \mathbb{Z}} f(aj)g(aj)\sigma^z_j .
\end{equation}
We therefore choose 
\begin{equation}
	Z_{\varphi}(a)Z_{\pi}(a) = \frac{a}{p'}
\end{equation}
to ensure that, in the limit $a\rightarrow 0$, we recover the usual canonical commutation relations.

We still have the freedom to choose $Z_{\varphi}(a)$ to ensure that the $n$-point correlation functions for $\widehat{\varphi}(x)$ exist. We detail the calculation for the $2$-point function, which, in turn, ensures the existence for the general case. Consider
\begin{equation}
	\begin{split}
	\langle \widetilde{\varphi}^{(a)}(f)\widetilde{\varphi}^{(a)}(g)\rangle &= Z_{\varphi}^2(a)\sum_{j,k\in \mathbb{Z}} f(aj)g(ak) \tr(\rho_{a}\sigma^x_{j}\sigma^x_{k}) \\
	 &= Z_{\varphi}^2(a)\sum_{j,k\in \mathbb{Z}} f(aj)g(ak) \tr(\rho_{a}\sigma^x_{0}\sigma^x_{k-j}) \\
	 &= Z_{\varphi}^2(a)\sum_{j,k'\in \mathbb{Z}} f(aj)g(a(k'+j)) \tr(\rho_{a}\sigma^x_{0}\sigma^x_{k'}) \\
	 &= Z_{\varphi}^2(a)\sum_{j,k'\in \mathbb{Z}} f(aj)g(a(k'+j)) C^{(a)}(ak') \\
	 &= Z_{\varphi}^2(a)\sum_{j\in \mathbb{Z}}\sum_{|k'| \le \lfloor 1/a \rfloor} f(aj)g(a(k'+j)) C^{(a)}(ak'),
	\end{split}
\end{equation}
where, for fixed $z=ak$,
\begin{equation}
	C^{(a)}(z) \equiv \tr(\rho_{a}\sigma^x_{0}\sigma^x_{\lfloor z/a\rfloor}) \underset{a\rightarrow 0}{\sim} a^{-2\log_2(\lambda_2)},
\end{equation}
so that $C(z) \equiv a^{2\log_2(\lambda_2)}C^{(a)}(z)$ has a well-defined limit.
To ensure that the limit exists we therefore choose
\begin{equation}
	Z_{\varphi}(a) = a^{\log_2(\lambda_2)+1} \equiv a^{\log_2(\sqrt{p}+\sqrt{1-p}) + \frac12}
\end{equation}
to give
\begin{equation}
	\begin{split}
	\langle \widetilde{\varphi}^{(a)}(f)\widetilde{\varphi}^{(a)}(g)\rangle &= a^{2\log_2(\lambda_2)+2}\sum_{j\in \mathbb{Z}}\sum_{|k'| \le \lfloor 1/a \rfloor} f(aj)g(a(k'+j)) C^{(a)}(ak') \\
	&= a^{2}\sum_{j\in \mathbb{Z}}\sum_{|k'| \le \lfloor 1/a \rfloor} f(aj)g(a(k'+j)) C(ak') 	\\
	&\underset{a\rightarrow 0}{\longrightarrow} \int_{-\infty}^{\infty} dx\, \int_{-1}^{1} dy\,  f(x)g(x+y) C(y).
	\end{split}
\end{equation}
Consequently,
\begin{equation}
	Z_\pi(a) = \frac{a^{-\log_2(\sqrt{p}+\sqrt{1-p}) + \frac12}}{p'}.
\end{equation}

\end{document}